\documentclass[preprint,pre,longbibliography,eqsecnum,nofootinbib]{revtex4-2} 

\usepackage{natbib}
\usepackage{graphicx}
\usepackage{subfigure}
\usepackage{amssymb}
\usepackage{amsfonts}
\usepackage{amsmath}
\usepackage{amsbsy}
\usepackage{mathrsfs} 
\usepackage{color}
\usepackage[colorlinks=true,citecolor=blue]{hyperref}
\usepackage{soul}
\usepackage{lineno}
\usepackage{mymacros}
\usepackage{graphicx}
\begin{document} 

\begin{abstract}
We study the onset of spontaneous dynamics in the follower force model of an active filament, wherein a slender elastic filament in a viscous liquid is clamped normal to a wall at one end and subjected to a tangential compressive force at the other. Clarke, Hwang and Keaveny (\textit{Phys. Rev. Fluids}, in press) have recently conducted a thorough investigation of this model using methods of computational dynamical systems; \textit{inter alia}, they show that the filament first loses stability via a supercritical double-Hopf bifurcation, with periodic `planar-beating' states (unstable) and `whirling' states (stable) simultaneously emerging at the critical follower-force value. We complement their numerical study by carrying out a weakly nonlinear analysis close to this unconventional bifurcation, using the method of multiple scales. The main outcome is an `amplitude equation' governing the slow modulation of small-magnitude oscillations of the filament in that regime. Analysis of this reduced-order model provides insights into the onset of spontaneous dynamics, including the creation of the nonlinear whirling states from particular superpositions of linear planar-beating modes as well as the selection of whirling over planar beating in three-dimensional scenarios. 
\end{abstract}

\title{Onset of spontaneous beating and whirling \\ in the follower force model of an active filament}
\author{Ory Schnitzer}
\affiliation{Department of Mathematics, Imperial College London, London SW7 2AZ, UK}

\maketitle
\newpage

\section{Introduction}
The translocation of molecular motors attached to biological filaments, such as microtubule or actin filaments, can give rise to rich dynamics and drive flows, as in ciliary beating or cytoplasmic streaming \citep{Shelley:16,Lauga:Book,Stein:21}. In particular, such active filaments may buckle under sufficiently strong compressive forces exerted by the motors. In that scenario, the tangential nature of the loading suggests an \emph{oscillatory} buckling instability leading to spontaneous dynamics.
 This contrasts the classical Euler buckling of a passive rod under a fixed load, where a monotonic instability leads to steady deformation. 

\citet{De:17} introduced a fundamental model consisting of a slender elastic filament clamped normal to a wall at one end and subjected to a tangential (compressive) `follower' force at the other, which is meant to phenomenologically represent the force exerted on the filament by a motor walking towards its tip. (They also considered a generalised model purporting to account for the opposite force applied by the motor on the fluid.) \citet{De:17} modelled the filament as an inertialess Kirchhoff rod \cite{Landau:BookE}, assumed that the filament motion is restricted to a plane and approximated the fluid-structure interaction according to `resistive force theory', a local anisotropic drag law valid at zero Reynolds number and to leading order in the slender-filament limit \cite{Lauga:Book}. By performing a linear-stability analysis, \citet{De:17} showed that their model filament undergoes symmetry breaking via a supercritical Hopf bifurcation; that is, the undeformed configuration of the filament becomes unstable at a critical follower-force magnitude, above which small perturbations grow in an oscillatory fashion. In that regime, \citet{De:17} demonstrated via initial-value simulations that the dynamics approach a limit cycle at late times, corresponding to the filament exhibiting  periodic `beating' oscillations. 

The planar-motion assumption was subsequently relaxed by \citet{Ling:18}, who considered a similar setup to that of \citet{De:17} but allowing for three-dimensional deformations and generalising from a localised follower force at the tip to arbitrarily distributed axial loads. Revisiting the case of a localised follower force, \citet{Ling:18} strikingly demonstrated through initial-value simulations that in a follower-force interval above the stability threshold the filament generally exhibits (at late times) \emph{non-planar} `whirling' oscillations --- periodic states for which a steadily rotating frame exists in which the deformed filament appears fixed; in that interval, the beating states observed by \citet{De:17} could only be attained by choosing the initial conditions to be precisely planar. At higher values of the follower force, i.e.~away from the instability threshold, the simulations of \citet{Ling:18} uncovered a secondary transition to a regime where planar beating is, in fact, preferred, as well as richer dynamics including chaotic-like phenomena. Distributing the load was shown to modify the nature of that secondary transition. 

To investigate the onset of spontaneous beating and whirling, \citet{Ling:18} carried out a linear-stability analysis allowing for general three-dimensional perturbations from the undeformed filament configuration. Similarly to \citet{De:17}, however, they found that the neutral modes at the instability threshold describe planar-beating oscillations of the filament. On that basis, \citet{Ling:18} concluded that \emph{`in the linear regime near the straight equilibrium, the filament undergoes planar deformations'}, thus leaving unexplained the emergence of whirling and its preference over planar beating. 

Partially motivated by this gap in understanding, \citet{Clarke:24} have recently carried out a thorough computational study of a generalised follower force model --- accounting for Stokes hydrodynamics fully rather than through resistive force theory. Upon revisiting the linear stability of the undeformed configuration, \citet{Clarke:24} made the following key observations. While the space of neutral linear modes at the instability threshold can indeed be spanned by planar-beating modes (similar in nature to those found by \citet{De:17} and \citet{Ling:18}), their superposition allowing for differences in phase and the plane of motion generally produces non-planar whirling-like modes --- the false conclusion of \citet{Ling:18} that the filament is restricted to planar deformations in the linear regime may accordingly be attributed to a misinterpretation of their linear analysis. In light of symmetry, the linear beating modes are identical up to a rotation about the axis of the undeformed configuration (as well as magnitude and phase differences), whereby the onset of instability occurs via a degenerate \emph{double-Hopf} bifurcation, where two identical pairs of complex-conjugate eigenvalues become unstable simultaneously. 

Employing methods of computational dynamical systems, \citet{Clarke:24} went beyond a linear-stability analysis of the undeformed configuration and initial-value simulations --- by computing and tracing the solution branches of periodic and quasi-periodic states, as well as determining the stability of periodic states via Floquet analysis. Their results reveal that two families of periodic states of the nonlinear problem, planar-beating states and whirling states, simultaneously emerge at the double-Hopf bifurcation. In an interval of the follower force above the threshold, the whirling states are stable while the planar-beating states are unstable under out-of-plane perturbations. Richer dynamics are uncovered at higher values of the follower force; in particular, the secondary transition to a regime of stable beating was shown to occur via a branch of quasi-periodic solutions connecting the whirling and beating solution branches. 

Complementary to the primarily numerical studies mentioned above, we shall here employ the method of multiple scales to carry out a weakly nonlinear analysis of the follower force model near its instability threshold. The main outcome is a reduced-order model in the form of a nonlinear `amplitude equation' governing the slow dynamics of small-magnitude oscillatory perturbations near the threshold, which is of higher dimension than the standard Stuart--Landau amplitude equation corresponding to a conventional (non-degenerate) Hopf instability \cite{Drazin:04}. We shall analyse and illustrate this reduced-order model in order to derive new insights into the onset of spontaneous dynamics in the follower force model. 
 
For the purpose of introducing our theoretical approach we wish to consider the follower force model in its most elementary form, though allowing for non-planar deformations which is clearly crucial in light of above discussion. We shall accordingly follow \citet{De:17} and \citet{Ling:18} by modelling the hydrodynamics via resistive force theory; follow \citet{De:17} and \citet{Clarke:24} by assuming a follower force localised at the tip; and follow \citet{Ling:18} and \citet{Clarke:24} by disregarding the biologically motivated entrainment flow included by \citet{De:17}. The preceding computational studies suggest that these approximations or modelling choices do not \emph{qualitatively} influence the dynamics of a single active filament in the vicinity of the instability threshold.
 
The paper proceeds as follows. In \S\ref{sec:formulation}, we formulate the problem. In \S\ref{sec:linear}, we revisit the linear stability of the undeformed configuration and the general linear approximation at the instability threshold. In \S\ref{sec:wnt}, we present, analyse and illustrate the reduced-order model, as well as validate the theory against numerical data provided by Dr Eric E.~Keaveny. The details of the weakly nonlinear analysis leading to the amplitude equation are retrospectively presented in \S\ref{sec:wna}. We conclude in \S\ref{sec:conclude} by discussing several future directions.

\section{Problem formulation}
\label{sec:formulation}
\subsection{Elastohydrodynamic model}
\label{ssec:model}
\begin{figure}[t!]
\begin{center}
\includegraphics[scale=0.4,trim={2cm 0.5cm 0 0}]{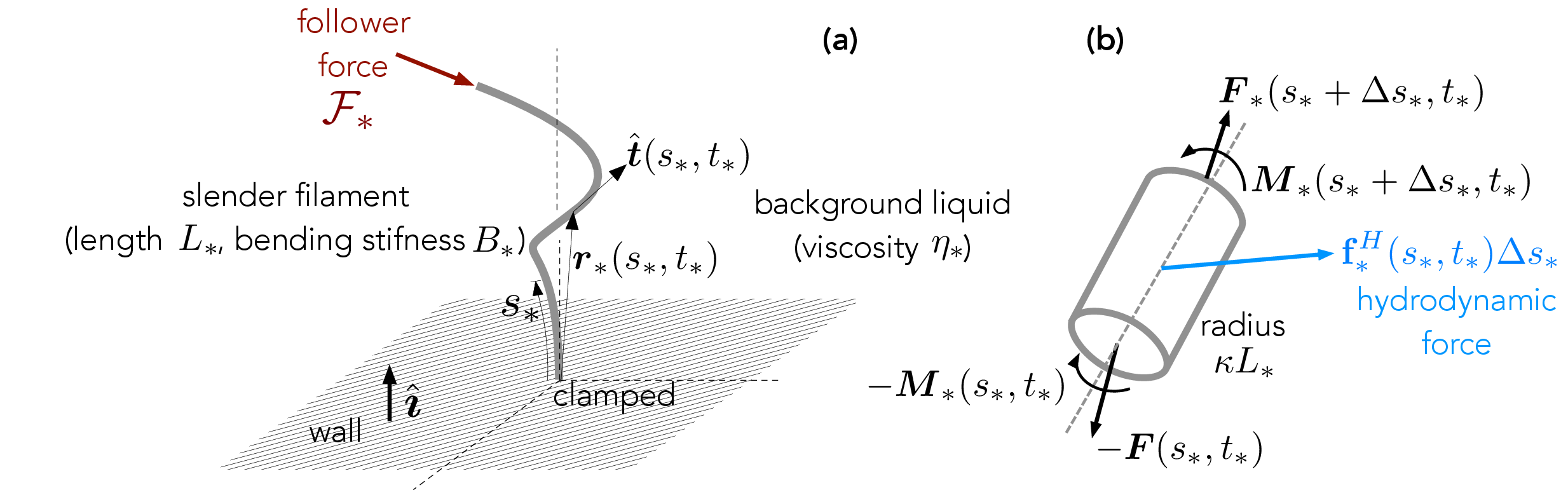}
\caption{(a) Dimensional schematic of the problem. (b) Forces and moments acting on a filament segment of small length $\Delta s_*$.}
\label{fig:schematic}
\end{center}
\end{figure}

As shown in Fig.~\ref{fig:schematic}, we consider an inextensible elastic filament of length $L_*$ and bending stiffness $B_*$ clamped normally to a flat wall at one end and subjected to a tangential compressive force of magnitude $\mathcal{F}_*$ at the other. The filament is immersed in a background liquid of viscosity $\eta_*$. We neglect inertia of the fluid and the filament and model the latter as a homogeneous Kirchhoff rod having a circular cross-section of radius $\kappa L_*$ \cite{Landau:BookE}, further assuming that it is sufficiently slender such that the hydrodynamics can be described by resistive force theory \cite{Graham:Book}. \emph{Throughout the paper, a subscript asterisk indicates a dimensional quantity; a superscript asterisk denotes complex conjugation; and $r$ and $i$ subscripts indicate the real and imaginary parts of a complex-valued quantity, respectively.} 

We represent the filament by its centreline $\br_*(s_*,t_*)$, where $s_*$ is the arc length measured from the clamped end and $t_*$ denotes the time. Since the filament is inextensible, the tangent vector $\bt=\partial\br_*/\partial s_*$ is constrained to be a unit vector. Let $\bF_*(s_*,t_*)$ and $\bM_*(s_*,t_*)$ be the cross-sectionally averaged internal (elastic) force and moment, respectively; with inertia neglected, they satisfy the equilibrium balances  
\refstepcounter{equation}
$$
\label{FM eqs}
\pd{\bF_*}{s_*}+\bbf_*^H=\bzero, \quad \pd{\bM_*}{s_*}+\bt\times\bF_*=\bzero,
\eqno{(\theequation \mathrm{a},\mathrm{b})}
$$ 
in which 
\begin{equation}\label{sbt dim}
\bbf_*^H=-\frac{4\pi\eta_*}{\ln(1/\kappa)}\left(\tI-\frac{1}{2}\bt\bt\right)\bcdot \pd{\br_*}{t_*}
\end{equation}
is the hydrodynamic force per unit length acting on the filament [see Fig.~\ref{fig:schematic}(b)]. The local anisotropic drag law \eqref{sbt dim} constitutes a leading-order asymptotic approximation of the hydrodynamics in the slender-filament limit $\kappa\to0$; under that approximation, known as resistive force theory, the interaction of the filament with itself and with the wall is neglected despite the \emph{relative} error associated with that approximation being only logarithmically small, scaling as $1/\ln(1/\kappa)$ \cite{Graham:Book}. 

The above governing equations are supplemented by initial conditions (which we do not specify at this stage), boundary conditions and a constitutive relation for the internal moment. The boundary conditions at the clamped end of the filament are
\refstepcounter{equation}
$$
\label{bc surface dim}
\br_*=\bzero, \quad \bt=\unit \quad \text{at} \quad s_*=0,
\eqno{(\theequation \mathrm{a},\!\mathrm{b})}
$$
where $\unit$ is the unit-vector normal to the wall pointing into the fluid and we choose the reference point for the position vector $\br_*$ to coincide with the clamping point. The boundary conditions at the other end of the filament are
\refstepcounter{equation}
$$
\label{bc tip dim}
\bF_*= -\mathcal{F}_*\bt, \quad \bM_*=\bzero \quad \text{at} \quad s_*=L_*,
\eqno{(\theequation \mathrm{a},\!\mathrm{b})}
$$
the first prescribing the follower force and the second stating that there is no external moment acting at the tip. Lastly, the constitutive law for the internal moment can be written as \cite{Landau:BookE}
\begin{equation}\label{constitutive dim}
\bM_*=B_*\bt\times\pd{\bt}{s_*},
\end{equation}
representing the resistance of the filament to bending, which is cross-sectionally isotropic given the assumption of a circular cross-section. In general, the constitutive relation includes an additional term representing the resistance of the filament to twisting. In the present scenario, however, where there are no external moments acting in the tangential direction and the bending is cross-sectionally isotropic, the twist term in the constitutive relation vanishes identically; this is verified in Appendix \ref{app:twist} following \citet{Landau:BookE}. The absence of twist in this scenario was also remarked by \citet{Ling:18}. In contrast, \citet{Clarke:24} must include twist as their model does not entail an approximation of the hydrodynamics in the thin-filament limit, whereby a finite viscous moment acts in the tangential direction.

\subsection{Dimensionless formulation}
\label{ssec:dimensionless}
It is convenient to adopt a dimensionless convention where lengths are normalised by $L_*$, forces by $B_*/L_*^2$, moments by $B_*/L_*$ and time by $4\pi\eta_*(L_*^4/B_*)/\ln(1/\kappa)$. \emph{Dimensionless fields are denoted similarly to their dimensional counterparts, only with the subscript asterisks omitted.} The problem thus consists of the partial differential equations 
\refstepcounter{equation}
$$
\label{eqs}
\pd{\br}{s}=\bt, \quad \pd{\bF}{s}-\left(\tI-\frac{1}{2}\bt\bt\right)\bcdot \pd{\br}{t} =\bzero, \quad \pd{\bM}{s}+\bt\times\bF=\bzero, \quad \bM=\bt\times\pd{\bt}{s};
\eqno{(\theequation \mathrm{a}\!-\!\mathrm{d})}
$$
the inextensibility constraint  
\begin{equation}\label{constraint}
\bt\bcdot\bt = 1;
\end{equation}
and the boundary conditions 
\refstepcounter{equation}
$$
\label{bcs wall}
\br=\bzero, \quad \bt=\unit \quad \text{at} \quad s=0 \eqno{(\theequation \mathrm{a},\mathrm{b})}
$$
and
\refstepcounter{equation}
$$
\label{bcs tip} 
\bF= -\mathcal{F}\bt, \quad \bM=\bzero \quad \text{at} \quad s=1,  \eqno{(\theequation \mathrm{a},\mathrm{b})}
$$
wherein the dimensionless parameter 
\begin{equation}\label{parameter}
\mathcal{F}=\frac{\mathcal{F}_*L_*^2}{B_*} 
\end{equation}
measures the magnitude of the follower force. The problem depends on this sole parameter, aside for the initial conditions which we do not specify at this stage. 

For all $\mathcal{F}$, the above problem possesses the steady solution 
\refstepcounter{equation}
$$
\label{base state}
\br=s\unit, \quad \bt = \unit, \quad \bF =-\mathcal{F}\unit, \quad \bM=\bzero, 
\eqno{(\theequation \mathrm{a}\!-\!\mathrm{d})}
$$
where the filament is undeformed (straight) and under uniform compression by the follower force. Our aim will be to study the nonlinear dynamics of the filament for $\mathcal{F}$ near the buckling threshold where this steady `base state' first becomes unstable.  

\section{Linear theory}
\label{sec:linear}
\subsection{Eigenvalue problem}
As a preliminary step, we must characterise the loss of stability of the base state. To this end, we assume small-magnitude perturbations of the form
\refstepcounter{equation}
$$
\label{linearisation}
\br-s\unit = e^{\lambda t}\tilde{\br} + \text{c.c.}  \quad \bt-\unit= e^{\lambda t}\tilde{\boldsymbol{t}} + \text{c.c.}, \quad \bF+\unit\mathcal{F}= e^{\lambda t}\tilde{\bF} + \text{c.c.}, \quad \bM= e^{\lambda t}\tilde{\bM} + \text{c.c.},
\eqno{(\theequation \mathrm{a}\!-\!\mathrm{d})}
$$
where $\lambda$ is a complex growth rate, the tilde-decorated vector fields are complex functions of $s$ and $\text{c.c.}$ stands for `complex conjugate'. Substituting \eqref{linearisation} into \eqref{eqs} yields, upon linearisation, the set of \emph{ordinary}  differential equations
\refstepcounter{equation}
$$
\label{eqs linear}
\tilde{\br}'=\tilde{\boldsymbol{t}}, \quad \tilde{\bF}'-\lambda\left(\tI-\frac{1}{2}\unit\unit\right)\bcdot \tilde{\br}=\bzero, \quad \tilde{\bM}'+\mathcal{F}\unit\times\tilde{\boldsymbol{t}}+\unit\times\tilde{\bF}=\bzero, \quad \tilde{\bM}=\unit\times\tilde{\boldsymbol{t}}',
\eqno{(\theequation \mathrm{a}\!-\!\mathrm{d})}
$$
where henceforth a prime denotes an ordinary derivative with respect to $s$ (e.g.~$\tilde{\br}'=\mathrm{d}\tilde{\br}/\mathrm{d}s$). 
Furthermore, linearisation of the constraint \eqref{constraint} gives 
\begin{equation}\label{constraint linear}
\unit\bcdot\tilde{\boldsymbol{t}} =0;
\end{equation} 
and linearisation of the boundary conditions \eqref{bcs wall} and \eqref{bcs tip} gives
\refstepcounter{equation}
$$
\label{bcs linear}
\tilde{\br}=\bzero, \quad \tilde{\boldsymbol{t}}=\bzero \quad \text{at} \quad s=0; \qquad \tilde{\bF}= -\mathcal{F}\tilde{\boldsymbol{t}}, \quad \tilde{\bM}=\bzero \quad \text{at} \quad s=1.
\eqno{(\theequation \mathrm{a}\!-\!\mathrm{d})} 
$$

It is readily seen that the tilde-decorated perturbations are all perpendicular to $\unit$. In particular, let $\tilde{\br}=\be_1\tilde{x}+\be_2\tilde{y}$, wherein $\tilde{x}(s)$ and $\tilde{y}(s)$ are complex-valued functions and $\be_1$ and $\be_2$ are unit vectors such that $\{\be_1,\be_2,\unit\}$ constitutes a right-handed orthogonal system. Substituting this Cartesian decomposition into \eqref{eqs linear} and \eqref{bcs linear}, we find that $\tilde{x}$ and $\tilde{y}$ are decoupled, separately satisfying the same eigenvalue problem. In terms of $\tilde{x}$, say, that problem consists of the ordinary differential equation 
\begin{equation}\label{ode}
\tilde{x}''''+\mathcal{F}\tilde{x}''+\lambda\tilde{x}=0
\end{equation}
and the boundary conditions
\refstepcounter{equation}
$$
\label{ode bcs}
\tilde{x}=0, \quad \tilde{x}'=0 \quad \text{at} \quad s=0; \qquad \tilde{x}''=0, \quad \tilde{x}'''=0 \quad \text{at} \quad s=1. 
\eqno{(\theequation \mathrm{a}\!-\!\mathrm{d})}
$$

\begin{figure}[t!]
\begin{center}
\includegraphics[scale=0.45,trim={1.5cm 1cm 0 0}]{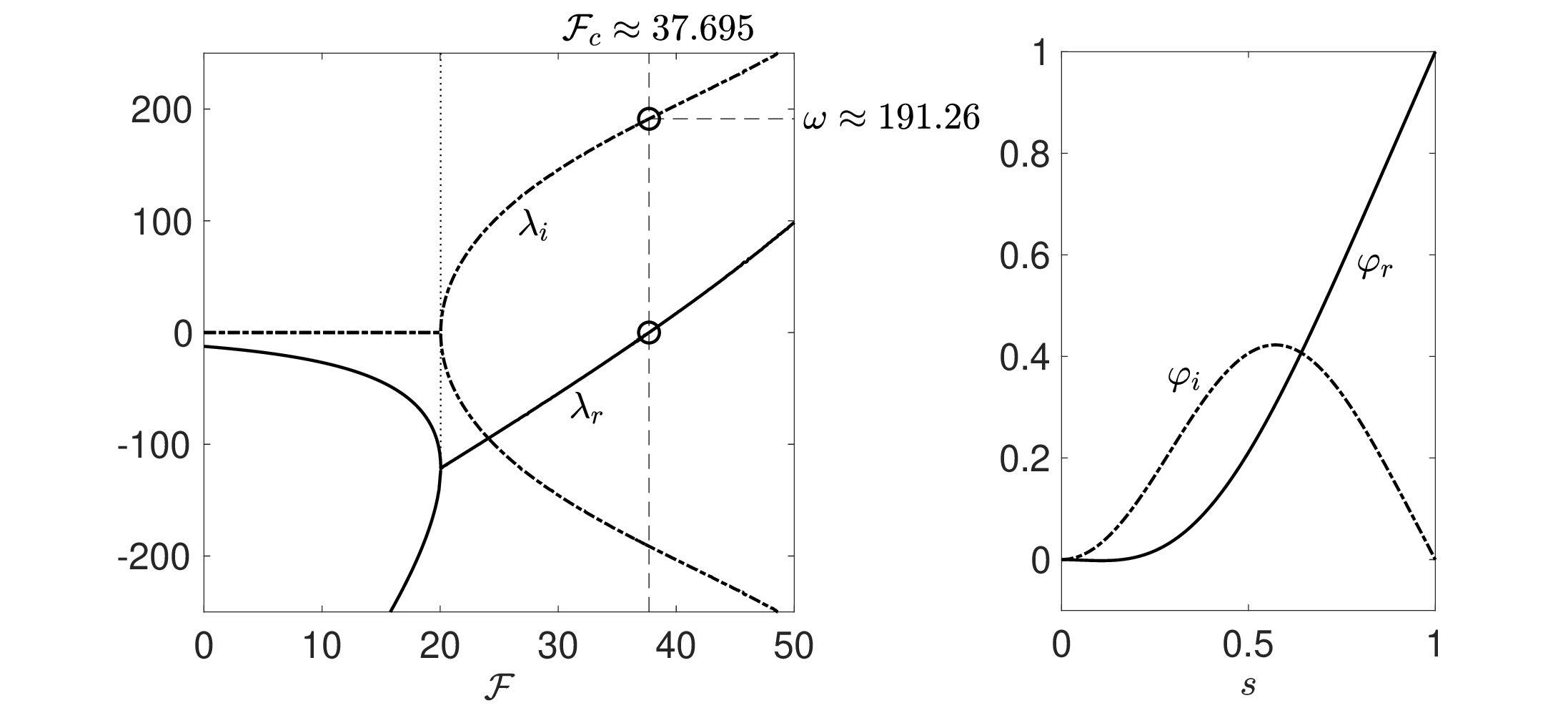}
\caption{(a) Complex growth rate $\lambda=\lambda_r+i\lambda_i$ as a function of $\mathcal{F}$, for the most unstable eigenfunctions of the eigenvalue problem consisting of \eqref{ode} and \eqref{ode bcs}. (b) The threshold eigenfunction $\varphi(s)=\varphi_r(s)+i\varphi_i(s)$ corresponding to the imaginary growth rate $\lambda=i\omega$.}
\label{fig:linear}
\end{center}
\end{figure}
The eigenvalue problem consisting of \eqref{ode} and \eqref{ode bcs} was first derived and solved by \citet{De:17} under the assumption of planar deformations. While the general solution to \eqref{ode} is readily expressed in closed form, calculating the eigenvalues $\lambda$ as a function of $\mathcal{F}$ must, in general, be done numerically. In agreement with \citet{De:17}, we thereby find that the base state is monotonically stable for $\mathcal{F}<\mathcal{F}_{o}\approx 20.051$, oscillatory stable for $\mathcal{F}_0<\mathcal{F}<\mathcal{F}_c\approx 37.695$ and oscillatory unstable for $\mathcal{F}>\mathcal{F}_c$; at the instability threshold, $\mathcal{F}=\mathcal{F}_c$, the complex-conjugate pair of neutral eigenvalues is $\lambda=\pm i\omega$, wherein $\omega \approx 191.26$ [see Fig.~\ref{fig:linear}(a)]. We denote by $\varphi(s)=\varphi_r(s)+i\varphi_i(s)$ the eigenfunction normalised to unity at $s=1$ that is associated with the threshold eigenvalue $i\omega$ [see Fig.~\ref{fig:linear}(b)]; $\varphi^*(s)$ is accordingly the similarly normalised eigenfunction associated with the conjugate eigenvalue $-i\omega$. 

\subsection{Double-Hopf bifurcation}
The eigenfunctions at the instability threshold describe periodic beating oscillations of the filament in the plane through the origin and normal to $\unit\times\be_1$, of arbitrary magnitude and phase. Since $\tilde{y}(s)$ satisfies the same eigenvalue problem as $\tilde{x}(s)$, there is clearly an independent pair of eigenfunctions representing similar beating oscillations rotated by $\pi/2$ about the undeformed configurations, i.e.~in the plane through the origin and normal to $\unit\times\be_2$. As first pointed out by \citet{Clarke:24}, the filament accordingly loses stability via a double-Hopf bifurcation, where two \emph{pairs} of complex-conjugate eigenvalues become unstable simultaneously; since the multiplicity here is induced by symmetry, the pairs are identical and have associated with them a full set of eigenfunctions. A key observation of \citet{Clarke:24} is that the linear superposition of beating modes allowing for differences in the phase and plane of motion generally produces non-planar whirling oscillations of the filament. 

\subsection{Linear approximation at the  threshold}
For our purposes, it will be convenient to represent the general linear whirling oscillations obtained by such superposition in terms of a complex-vector\footnote{Let $\ba=\ba_r+i\ba_i$ and $\bb=\bb_r+i\bb_i$ be arbitrary complex vectors in three-dimensional space. The dot and cross products are generalised from the case of real vectors as $\ba\bcdot\bb=\ba_r\bcdot\bb_r-\ba_i\bcdot\bb_i+i(\ba_r\bcdot\bb_i+\ba_i\bcdot\bb_r)$ and $\ba\times\bb=\ba_r\times\bb_r-\ba_i\times\bb_i+i(\ba_r\times\bb_i+\ba_i\times\bb_r)$. An appropriate inner product is $\langle \ba,\bb\rangle=\ba^*\bcdot\bb$, which differs from the dot product. The magnitude of a complex vector $\ba$ is accordingly defined as $|\ba|=\sqrt{\ba^*\bcdot\ba}=\sqrt{|\ba_r|^2+|\ba_i|^2}$.} amplitude $\tilde{\bA}$ parallel to the wall. Noting from \eqref{eqs linear} that an eigenfunction $\tilde{\br}=\be_1\varphi$ is associated with perturbations $\tilde{\boldsymbol{t}}=\be_1\varphi'$, $\tilde{\bF}=-\be_1(\varphi'''+\mathcal{F}_c\varphi')$ and $\tilde{\bM}=\unit\times\be_1\varphi''$, a general linear approximation at the threshold $\mathcal{F}=\mathcal{F}_c$ can be expressed as [cf.~\eqref{linearisation}]
\begin{subequations}
 \label{general linear} 
 \begin{eqnarray}
 \br-s\unit &=& e^{i\omega t}\tilde{\bA}\varphi + \text{c.c.},\\
 \bt-\unit &=& e^{i\omega t}\tilde{\bA}\varphi' + \text{c.c.},\\ 
 \bF+\unit\mathcal{F}_c &=& -e^{i\omega t}\tilde{\bA}(\varphi'''+\mathcal{F}_c\varphi') + \text{c.c.}, \\
 \bM &=& e^{i\omega t}\unit\times\tilde{\bA}\varphi'' + \text{c.c.},
\end{eqnarray}
\end{subequations}
where we have ignored decaying stable modes. 
Given the normalization $\varphi(1)=1$, the corresponding tip trajectory is 
\begin{equation}\label{tip motion}
\br(1,t) -\unit = \tilde{\bA}e^{i\omega t}+ \text{c.c}=2\tilde{\bA}_r\cos\omega t-2\tilde{\bA}_i\sin\omega t,
\end{equation}
which traces a general ellipse parallel to the wall and centered about the nominal tip position, the vector $\mathrm{Im}(\tilde{\bA}\times\tilde{\bA}^*)$ indicating the direction of motion according to the right-hand rule. It is easy to see that for $\tilde{\bA}\times\tilde{\bA}^*=\bzero$, the ellipse degenerates to a line (parallel to $\tilde{\bA}_r$ and $\tilde{\bA}_i$), of length $4|\tilde{\bA}|$. For $\tilde{\bA}\bcdot\tilde{\bA}=0$, the ellipse constitutes a circle of radius $\sqrt{2}|\tilde{\bA}|$. 

The properties of a general tip ellipse can be derived following \citet{Lindell:83}. In particular, the lengths of the semi-major and semi-minor axes can be expressed as 
\begin{equation}\label{axes}
\sqrt{|\tilde{\bA}|^2+|\tilde{\bA}\times\tilde{\bA}^*|}\pm\sqrt{|\tilde{\bA}|^2-|\tilde{\bA}\times\tilde{\bA}^*|},
\end{equation}
respectively, giving the ellipse area as $2\pi|\tilde{\bA}\times\tilde{\bA}^*|$. [The above-mentioned condition for a circular orbit, $\tilde{\bA}\bcdot\tilde{\bA}=0$, can be inferred from \eqref{axes} by noting the identity $|\tilde{\bA}\bcdot\tilde{\bA}|^2=|\tilde{\bA}|^4-|\tilde{\bA}\times\tilde{\bA}^*|^2$.] Furthermore, for non-circular ellipses the real and imaginary parts of $\tilde{\bA}/(\tilde{\bA}\bcdot\tilde{\bA})^{1/2}$ point in the directions of the major and minor axes, respectively (both the forward and backward axis directions, given the multiplicity of the square-root function).\footnote{The formulae for the elliptical tip trajectory  can alternatively be expressed in terms of the real vectors $\tilde{\bA}_r$ and $\tilde{\bA}_i$, by noting the identities $\tilde{\bA}\times\tilde{\bA}^*=-2i\tilde{\bA}_r\times\tilde{\bA}_i$ and $\tilde{\bA}\bcdot\tilde{\bA}=|\tilde{\bA}_r|^2-|\tilde{\bA}_i|^2+2i\tilde{\bA}_r\bcdot\tilde{\bA}_i$. Hence, the ellipse is a line if $\tilde{\bA}_r\times\tilde{\bA}_i=\bzero$; it is a circle if both $\tilde{\bA}_r\bcdot\tilde{\bA}_i=0$ and $|\tilde{\bA}_r|=|\tilde{\bA}_i|$.} 

The trajectory $c\tilde{\bA}e^{i\omega t}+\text{c.c.}$, wherein $c$ is a complex scalar, is identical to the elliptical tip trajectory \eqref{tip motion} up to a dilation $|c|$ and phase $\angle c$. It thus follows from (\ref{general linear}a) that, in fact, all points along the filament centreline carry out geometrically similar elliptical trajectories, obtained from the tip trajectory by a dilation $|\varphi(s)|$ and phase $\angle\varphi(s)$.  From (\ref{general linear}b-d), the tangent unit vector, force and moment perturbations along the filament can be analogously constructed by appropriate scale and phase modulations of the elliptical tip trajectory (as well as a $\pi/2$ rotation in the case of the moment perturbation). 

It can be confirmed from the above formulae that in the special case of circular whirling, $\tilde{\bA}\bcdot\tilde{\bA}=0$, any given point along the filament performs a steady circular motion parallel to the wall. Furthermore, in that case the geometry of the deformed filament and the force and moment distributions along it appear fixed in a frame of reference rotating about the undeformed configuration at a steady angular frequency $\omega$. 

\section{Weakly nonlinear theory}
\label{sec:wnt}
\subsection{Separation of time scales near the instability threshold}\label{ssec:scalings}
According to the linear theory, at the instability threshold the filament approaches with time periodic beating/whirling oscillations of the form \eqref{general linear}, with the complex-vector amplitude $\tilde{\bA}$, assumed small in magnitude, fixed by the initial conditions. In fact, the fate of small-magnitude oscillations at the instability threshold is governed by weak nonlinear effects which become important at  late times --- invalidating the linear theory! Specifically, nonlinear terms formed of cubic products of the linear approximation (or its derivatives) may resonantly excite the neutral linear modes, thereby producing perturbations growing like $t|\tilde{\bA}|^3$. Over $O(1/|\tilde{\bA}|^{2})$ times --- long compared to the natural period $2\pi/\omega$ --- such nonlinear perturbations would become comparable to the linear approximation and may accordingly influence the dynamics. 

On the other hand, the linear theory predicts that small-magnitude perturbations exponentially decay for $\mathcal{F}<\mathcal{F}_c$, or exponentially grow, in general, for $\mathcal{F}>\mathcal{F}_c$. As $\mathcal{F}\to\mathcal{F}_c$, the growth rate $\lambda_r$ of the most unstable linear modes vanishes like $\mathcal{F}-\mathcal{F}_c$ (see Fig.~\ref{fig:linear}a), suggesting that following a  transient (during which more stable modes decay) the filament exhibits oscillations of the form \eqref{general linear} --- modulated over long $O(1/|\mathcal{F}-\mathcal{F}_c|)$ times. We accordingly identify the distinguished scaling $|\tilde{\bA}|=\text{ord}(|\mathcal{F}-\mathcal{F}_c|^{1/2})$ --- conventional to symmetry-breaking bifurcations --- such that linear and nonlinear effects both have a leading-order effect. For larger amplitudes, or at the threshold, the dynamics are dominated by nonlinear effects; for smaller amplitudes, they are dominated by the linear dynamics.

The time-scale separation at and near the instability threshold can be exploited towards systematically deriving a reduced-order nonlinear model governing the dynamics of small-magnitude oscillations in this regime. This is carried out in \S\ref{sec:wna} by means of a weakly nonlinear analysis of the governing equations employing the method of multiple scales; while conceptually standard, the derivation is presented fully so as to highlight some technical details peculiar to the present setup and, more importantly, to facilitate future extensions to the theory as discussed in \S\ref{sec:conclude}. In the present section, we simply quote the reduced-order model and subsequently analyse and illustrate it in various scenarios. 

\subsection{Amplitude equation}\label{ssec:amplitude}
The reduced-order model consists of the linear approximation \eqref{general linear}, now understood to hold in a vicinity of the instability threshold and with $\tilde{\bA}$ evolving with time according to the amplitude equation
\begin{equation}\label{amplitude primitive}
\frac{\mathrm{d}\tilde{\bA}}{\mathrm{d}t}=\alpha\tilde{\bA}^*\tilde{\bA}\bcdot\tilde{\bA}+\beta\tilde{\bA}{\tilde{\bA}}^*\bcdot\tilde{\bA}+(\mathcal{F}-\mathcal{F}_c)\gamma\tilde{\bA},
\end{equation}
in which the complex coefficients $\alpha$, $\beta$ and $\gamma$ are provided in Appendix \ref{app:coefficients} as quadratures of the eigenfunction $\varphi(s)$, yielding the numerical values  
\refstepcounter{equation}
$$\label{coefficients numbers}
\alpha \approx -451.038 -i327.599,\quad \beta\approx -517.974+i353.952, \quad \gamma\approx 7.34480+i5.34285.
\eqno{(\theequation \mathrm{a}\!-\!\mathrm{c})}
$$
In accordance with the above scaling remarks, this reduced-order model holds in the limit of small-magnitude perturbations $|\tilde{\bA}|\ll1$, with $\mathcal{F}-\mathcal{F}_c=O(|\tilde{\bA}|^2)$. Furthermore, we note that the first two (nonlinear) terms on the right-hand side of \eqref{amplitude primitive} imply the long time scale $1/|\tilde{\bA}|^2$, whereas the third (linear) term implies the long time scale $1/|\mathcal{F}-\mathcal{F}_c|$.  

We prefer to work with rescaled, order-unity quantities. Thus, consistently with the weakly nonlinear analysis in \S\ref{sec:wna}, we define 
\begin{equation}\label{F shift}
\mathcal{F}=\mathcal{F}_c+\epsilon\chi,
\end{equation}
with $0<\epsilon\ll1$ a small positive parameter and $\chi$ real, and introduce the rescalings
\begin{equation}\label{rescalings}
\tilde{\bA}(t)=\epsilon^{1/2}\bA(T), 
\end{equation}
wherein
\begin{equation}
T=\epsilon t
\end{equation}
is a slow time coordinate. 
The amplitude equation \eqref{amplitude primitive} then reads as
\begin{equation}
\label{amplitude}
\frac{\mathrm{d}{\bA}}{\mathrm{d}T}=\alpha{\bA}^*{\bA}\bcdot{\bA}+\beta{\bA}{{\bA}}^*\bcdot{\bA}+\chi\gamma{\bA}.
\end{equation}
While only the product $\epsilon\chi$ is meaningful, the added freedom is convenient for jointly scaling the cases where the follower-force magnitude is at or near the instability threshold; in the latter case, the freedom can be removed by setting $\chi=\mathrm{sgn}(\mathcal{F}-\mathcal{F}_c)$.  

It is useful to note the following three relations:
\begin{subequations}
 \label{relations}
 \begin{eqnarray}
  \frac{\mathrm{d}}{\mathrm{d}T}(\bA^*\times\bA)&=&2(\beta_r|\bA|^2+\chi\gamma_r)\bA^*\times\bA,\\
   \frac{\mathrm{d}}{\mathrm{d}T}(\bA\bcdot\bA)&=& 2\left[(\alpha+\beta)|\bA|^2+\chi\gamma\right]\bA\bcdot\bA,\\
 \frac{\mathrm{d}}{\mathrm{d}T}|\bA|^2&=&2(\alpha_r+\beta_r)|\bA|^4-2\alpha_r|\bA^*\times\bA|^2+2\chi\gamma_r|\bA|^2,
 \end{eqnarray}
 \end{subequations}
 which readily follow from \eqref{amplitude}.\footnote{We obtain (\ref{relations}a) by subtracting the cross product of $\bA^*$ and \eqref{amplitude} from the conjugate of that product. We obtain (\ref{relations}b) by considering the dot product of $\bA$ and \eqref{amplitude}, noting the identity $\bA\bcdot\frac{\mathrm{d}\bA}{\mathrm{d}T}=\frac{1}{2}\frac{\mathrm{d}}{\mathrm{d}T}(\bA\bcdot\bA)$. We obtain (\ref{relations}c) by adding the dot product of $\bA^*$ and \eqref{amplitude} to the conjugate of that product.} 
In particular, (\ref{relations}a) shows that if the fast-scale tip orbit is initially a line (corresponding to planar motion of the filament), then it remains so at all later slow times.  Similarly, (\ref{relations}b) shows that if the fast-scale tip orbit is initially circular then it remains so at all later slow times. 

If the filament is confined to a plane (by the initial conditions), the complex-vector amplitude can be written $\bA(T)=\be A(T)$, where $A(T)$ is a complex-scalar amplitude and $\be$ a unit vector parallel to the wall; the filament in that case performs planar-beating oscillations with $A(T)$ determining their magnitude and phase. In that restricted scenario, the amplitude equation \eqref{amplitude} reduces to 
\begin{equation}\label{planar amplitude}
\frac{\mathrm{d}A}{\mathrm{d}T}=(\alpha+\beta)|A|^2A+\chi\gamma A,
\end{equation}
which has the form of the Stuart--Landau amplitude equation corresponding to a conventional (non-degenerate) Hopf instability \cite{Drazin:04}. 

We proceed in \S\ref{ssec:steady} to find solutions of the amplitude equation \eqref{amplitude} representing steady and periodic states; we also compare these solutions against numerical data provided by Dr Eric E.~Keaveny. In \S\ref{ssec:stability_stationary}-\ref{ssec:nonlinear_stab}, we analyse the stability of these states within the framework of the weakly nonlinear theory. In \S\ref{ssec:illustrate}, we present simulations of \eqref{amplitude}.

\subsection{Periodic states}\label{ssec:steady}
We consider `quasi-steady' solutions of the amplitude equation \eqref{amplitude} in the form 
\begin{equation}\label{quasi form}
\bA=e^{i\nu T}\bar{\bA},
\end{equation}
wherein $\bar{\bA}$ is a time-independent complex vector parallel to the wall and $\nu$ a real scalar. Such solutions represent periodic states of the filament at a corrected angular frequency $\omega+\nu\chi^{-1}(\mathcal{F}-\mathcal{F}_c)$. From \eqref{amplitude}, the reduced amplitude $\bar{\bA}$ satisfies 
\begin{equation}\label{steady amplitude}
\alpha\bar{\bA}^*\bar{\bA}\bcdot\bar{\bA}+\beta\bar{\bA}\bar{\bA}^*\bcdot \bar{\bA}+(\chi\gamma-i\nu)\bar{\bA}=\bzero. 
\end{equation}
The trivial solution $\bar{\bA}=\bzero$ corresponds to the undeformed state of the filament. By crossing \eqref{steady amplitude} with $\bar{\bA}$, we find that nontrivial quasi-steady must satisfy either $\bar{\bA}\times\bar{\bA}^*=\bzero$, corresponding to planar beating, or $\bar{\bA}\bcdot\bar{\bA}=0$, corresponding to circular whirling. 

\subsubsection{Planar-beating states}\label{sssec:beating}
In the planar-beating case, we can write 
\begin{equation}\label{beating form}
\bar{\bA}=\bar{A}\be, 
\end{equation} 
wherein $\be$ is a unit vector parallel to the wall and $\bar{A}$ is a non-vanishing complex scalar. Representing the latter in polar form,
\begin{equation}\label{beating form 2}
\bar{A}=e^{i\vartheta}|\bar{A}|,
\end{equation} 
wherein $\vartheta$ represents a real phase, 
we find from \eqref{steady amplitude} that the  magnitude $|\bar{A}|$ satisfies
\begin{equation}\label{beating eq}
(\alpha+\beta)|\bar{A}|^2+\chi\gamma-i\nu=0,
\end{equation}
whereas the direction $\be$ and phase $\vartheta$ are  arbitrary. 
Considering the real and imaginary parts of \eqref{beating eq}, we readily find 
\refstepcounter{equation}
\label{beating sol}
$$
|\bar{A}|^2=-\frac{\chi\gamma_r}{\alpha_r+\beta_r}, \quad \nu=\chi\gamma_i-\chi\gamma_r\frac{\alpha_i+\beta_i}{\alpha_r+\beta_r}.
\eqno{(\theequation \mathrm{a},\mathrm{b})}
$$
Given the coefficient values provided by \eqref{coefficients numbers}, the right-hand side of (\ref{beating sol}a) implies that the planar-beating states exist only in the case $\chi>0$, namely when the follower-force magnitude is above the instability threshold. Setting $\chi=1$, without loss of generality, the corresponding tip trajectories follow from \eqref{tip motion}, using \eqref{rescalings}, \eqref{quasi form}, \eqref{beating form} and \eqref{beating form 2}, as
\begin{equation}\label{beating tip}
\br(1,t) -\unit =  2|\bar{A}|(\mathcal{F}-\mathcal{F}_c)^{1/2}  \cos[(\omega +|\mathcal{F}-\mathcal{F}_c| \nu) t+\vartheta]\be,
\end{equation}
where the rescaled tip displacement $2|\bar{A}|\doteq0.17412$ and angular-frequency correction $\nu\doteq5.5426$ are calculated using \eqref{coefficients numbers} and \eqref{beating sol}.

\begin{figure}[t!]
\begin{center}
\includegraphics[scale=0.45,trim={1.5cm 1cm 0 0}]{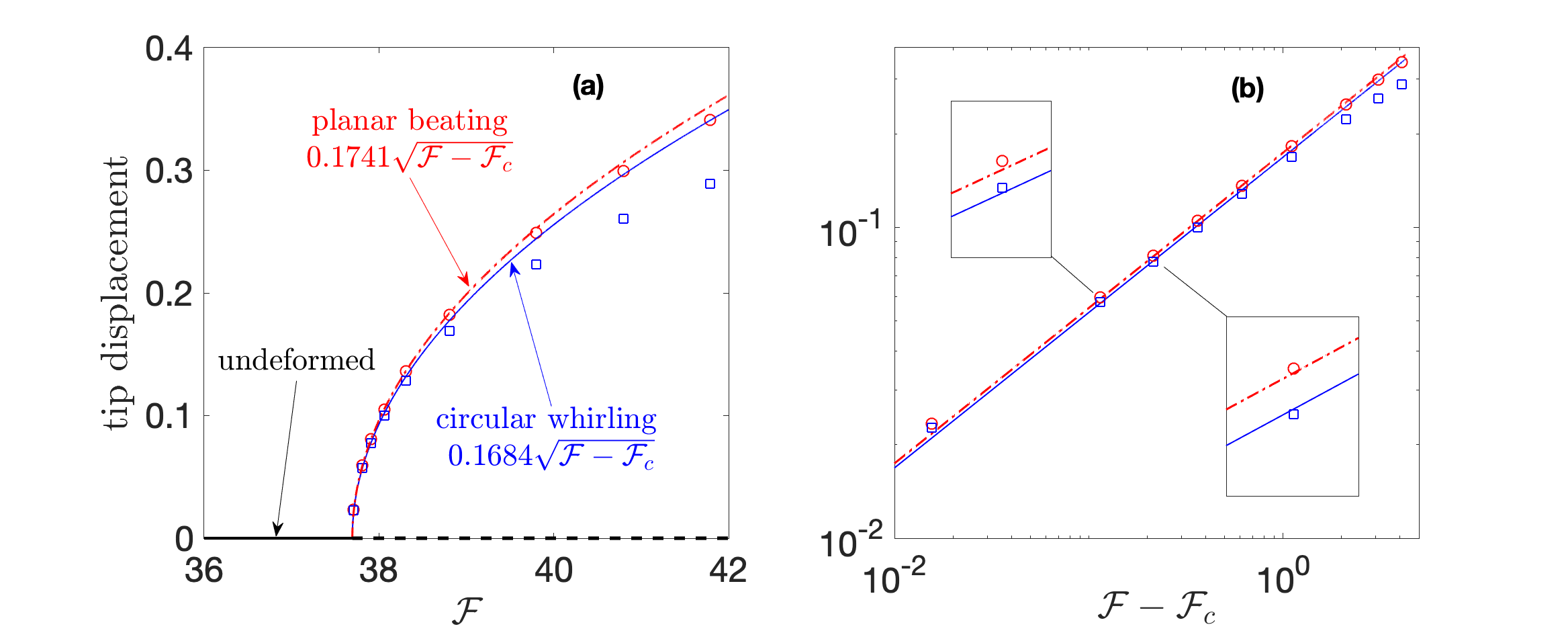}
\caption{(a) Bifurcation diagram showing maximal dip displacement (normalised by filament length) against the dimensionless follower-force magnitude $\mathcal{F}$, for the (i) undeformed steady state; (ii) planar-beating periodic states; and (iii) circular-whirling periodic states. The curves depict the predictions of the weakly nonlinear theory (see \S\ref{sec:wnt}): the undeformed state is stable preceding the bifurcation (solid black line) and unstable following it (dashed black line); the planar-beating states are longitudinally stable, up to phase shifts, but transversally unstable (dash-dotted red curve); and the circular-whirling states are stable up to phase shifts (solid blue curve). The symbols depict numerical data provided by Dr Eric. E.~Keaveny (see \S\ref{ssec:validation}). (b) Zoomed-in comparison between the theory and numerical data depicted on a log-log scale.} 
\label{fig:states}
\end{center}
\end{figure}
\subsubsection{Circular-whirling states}\label{sssec:cw}
In the circular-whirling case, $\bar{\bA}\bcdot\bar{\bA}=0$, the real vectors $\bA_r$ and $\bA_i$ are orthogonal and of equal magnitude. We accordingly write
\begin{equation}\label{cw form}
\bar{\bA}=\frac{1}{\sqrt{2}}|\bar{\bA}|(\be \pm i\unit\times\be),
\end{equation}
where the magnitude $|\bar{\bA}|\ne0$; the plus/minus sign indicates clockwise/counterclockwise motion (viewed from above the wall), respectively; and ${\be}$ is a unit-vector parallel to the wall which represents a phase. From \eqref{steady amplitude}, the magnitude $|\bar{\bA}|\ne0$ satisfies
\begin{equation}\label{cw eq}
\beta|\bar{\bA}|^2+\gamma\chi-i\nu=0,
\end{equation}
whereas the direction of motion and the phase unit-vector $\be$ are arbitrary. Considering the real and imaginary parts of \eqref{cw eq}, we readily find \refstepcounter{equation}
\label{cw sol}
$$
|\bar{\bA}|^2 = -\frac{\chi\gamma_r}{\beta_r}, \quad \nu = \chi\gamma_i-\chi\gamma_r\frac{\beta_i}{\beta_r}. 
\eqno{(\theequation \mathrm{a},\mathrm{b})}
$$
Similarly to the planar-beating states, we see from the right-hand side of (\ref{cw sol}a) and the coefficient values provided by \eqref{coefficients numbers} that the circular-whirling states exist only in the case $\chi>0$, similarly to the planar-beating states. Setting $\chi=1$, without loss of generality, the corresponding tip trajectories follow from \eqref{tip motion}, using \eqref{rescalings}, \eqref{quasi form} and \eqref{cw form}, as 
\begin{multline}\label{cw tip}
\br(1,t) -\unit =  \sqrt{2}|\bar{\bA}|(\mathcal{F}-\mathcal{F}_c)^{1/2}\left\{\be\cos\left[(\omega +|\mathcal{F}-\mathcal{F}_c| \nu) t\right] \right. \\ \left. \mp \unit\times\be \sin\left[(\omega+|\mathcal{F}-\mathcal{F}_c|\nu)t\right] \right\},
\end{multline}
where the rescaled tip radius $\sqrt{2}|\bar{\bA}|\doteq0.16840$ and angular-frequency correction $\nu\doteq10.362$ are calculated using \eqref{coefficients numbers} and \eqref{cw sol}.

\subsubsection{Bifurcation diagram and comparison with numerical solutions of the full problem}\label{ssec:validation}
In Fig.~\ref{fig:states}, we plot maximal tip displacement vs.~dimensionless follower force for the steady and periodic states predicted by the weakly nonlinear theory. As evident in \eqref{beating tip} and \eqref{cw tip}, the maximal tip displacement $\propto \sqrt{\mathcal{F}-\mathcal{F}_c}$ for both the planar-beating and circular-whirling periodic states, the constant of proportionality being slightly larger for the planar-beating states. (We note that this square-root scaling near the threshold was hypothesised as well as numerically demonstrated by \citet{Clarke:24}.) The stability characteristics indicated in the bifurcation diagram will be derived and discussed in the following subsections. 

With the purpose of validating the present weakly nonlinear theory, Dr Eric E.~Keaveny has conducted simulations based on the numerical methodology of \citet{Schoeller:21}, adopting the same formulation as \citet{Clarke:24} only with the full Stokes hydrodynamics included in their study simplified to resistive force theory ---  consistently with the present formulation [cf.~\eqref{sbt dim}]. Fig.~\ref{fig:states} shows remarkable agreement, in terms of the maximal tip displacements, between these simulations and the theoretical predictions.

\subsection{Linear stability of the undeformed state} \label{ssec:stability_stationary}
The stability characteristics of the undeformed state, as well as the planar-beating and circular-whirling periodic states, are readily determined within the framework of the weakly nonlinear theory. We begin by revisiting the stability of the undeformed state, which has already been analysed exactly in \S\ref{sec:linear}. To this end, we consider small perturbations of the complex-vector amplitude, 
\begin{equation}\label{linearisation stationary}
\bA(T)=\ba(T), \quad |\ba|\ll1.
\end{equation}
Linearisation of the amplitude equation \eqref{amplitude} in the case $\bar{\bA}=\bzero$ gives
\begin{equation}\label{stability stationary lin}
\frac{\mathrm{d}\ba}{\mathrm{d}T}=\chi\gamma\ba,
\end{equation}
which possesses the general solution 
\begin{equation}\label{stability stationary lin sol}
\ba(T)=e^{\chi\gamma T}\ba(0).
\end{equation}
In this linear approximation, the elliptical tip orbit at slow time $T$ is obtained from the initial orbit defined by $\ba(0)$ via a dilation by $\exp(\chi\gamma_r T)$. Since $\gamma_r>0$, perturbations grows for $\chi>0$ and decay for $\chi<0$. The case $\chi=0$, where nonlinear terms determine the stability of the undeformed state, will be considered in \S\ref{ssec:nonlinear_stab}.  

It is instructive to repeat the above stability analysis, this time starting from the real dynamical system implied by  the linearised amplitude equation \eqref{stability stationary lin}, 
\begin{equation}
\label{stability stationary sys}
\frac{\mathrm{d}}{\mathrm{d}T}\left(\begin{array}{c}\ba_r \\ \ba_i \end{array}\right) = \chi\left(\begin{array}{cc} \gamma_r & -\gamma_i \\\gamma_i & \gamma_r\end{array}\right)\left(\begin{array}{c}\ba_r \\ \ba_i \end{array}\right).
\end{equation}
As $\ba_r$ and $\ba_i$ are real vectors parallel to the wall, this system is four-dimensional. By substituting into \eqref{stability stationary sys} the modal ansatz 
\begin{equation}\label{stability stationary modal}
\left(\begin{array}{c}\ba_r \\ \ba_i \end{array}\right)=e^{\sigma T}\left(\begin{array}{c}\tilde{\ba}_r \\ \tilde{\ba}_i \end{array}\right)+\text{c.c.},
\end{equation}
wherein $\sigma$ represents a rescaled growth rate, we find the homogeneous system
\begin{equation}
\label{stability stationary sys homo}
 \left(\begin{array}{cc} \chi\gamma_r-\sigma & -\chi\gamma_i \\\chi\gamma_i & \chi\gamma_r-\sigma\end{array}\right)\left(\begin{array}{c}\tilde{\ba}_r \\ \tilde{\ba}_i\end{array}\right)=\left(\begin{array}{c}\bzero \\ \bzero\end{array}\right).
\end{equation}
The solution space is spanned by the eigenvectors 
\refstepcounter{equation}
$$
\label{stationary stability eigenvectors}
\left(\begin{array}{c}\tilde{\ba}_r \\ \tilde{\ba}_i \end{array}\right)\ub{1,2}=\left(\begin{array}{c}\be_{1} \\ \mp i\be_{1} \end{array}\right), \quad  \left(\begin{array}{c}\tilde{\ba}_r \\ \tilde{\ba}_i \end{array}\right)\ub{3,4}=\left(\begin{array}{c}\be_{2} \\ \mp\be_{2} \end{array}\right),
\eqno{(\theequation \mathrm{a},\mathrm{b})}
$$
$\be_1$ and $\be_2$ being a pair of orthogonal unit vectors parallel to the wall, with corresponding growth-rate eigenvalues
\refstepcounter{equation}
$$
\label{stationary eigenvalues}
\sigma\ub{1,3}=\chi\gamma, \quad \sigma\ub{2,4}=\chi\gamma^*.
\eqno{(\theequation \mathrm{a},\mathrm{b})}
$$
In accordance with the stability analysis in \S\ref{sec:linear}, we find two identical pairs of complex-conjugate growth rates whose real part is positive for $\chi>0$ and negative for $\chi<0$. The eigenvector pairs (\ref{stationary eigenvalues}a) and (\ref{stationary eigenvalues}b) correspond to planar beating along the $\be_1$ and $\be_2$ directions, respectively. The linearised whirling motion \eqref{stability stationary lin sol} is readily retrieved by general superposition of solutions in the form \eqref{stability stationary modal}.

\subsection{Linear stability of the periodic states} \label{ssec:stability_nontrivial}
We next consider the stability of the planar-beating and circular-whirling states. In these cases, it is convenient to write the perturbation as 
\begin{equation}\label{stability nontrivial perturbation}
e^{-i\nu T}\bA(T) = \bar{\bA}+\ba(T), \quad |\ba|\ll1,
\end{equation}
wherein the steady complex-vector amplitude $\bar{\bA}$ and angular-frequency correction $\nu$ are defined as in \eqref{quasi form}. Since these states exist only in the case $\chi>0$, we set $\chi=1$, without loss of generality,  for the remainder of this subsection. Linearisation of the amplitude equation \eqref{amplitude} then gives, upon eliminating the quasi-steady relation \eqref{steady amplitude}, 
\begin{equation}\label{linear amplitude}
\frac{\mathrm{d}\ba}{\mathrm{d}T}=2\alpha{\bar{\bA}}^*\bar{\bA}\bcdot \ba+\alpha\ba^*\bar{\bA}\bcdot \bar{\bA}+\beta\bar{\bA}\bar{\bA}\bcdot \ba^*+\beta\bar{\bA}\ba\bcdot{\bar{\bA}}^*+(\gamma-i\nu+\beta|\bar{\bA}|^2)\ba.
\end{equation}

\subsubsection{Linear stability of the planar-beating states}\label{sssec:beating_stability}
For the planar-beating states (\S\ref{sssec:beating}), the form of $\bar{\bA}$ is given by \eqref{beating form} and \eqref{beating form 2}, which feature the unit vector $\be$ (indicating the plane of beating) and phase $\vartheta$. Decomposing the perturbation into components parallel and perpendicular to the beating, $\ba=a_{\parallel}\be+a_{\perp}\unit\times\be$, the linearised amplitude equation \eqref{linear amplitude} yields, upon eliminating $\nu$ using the steady relation \eqref{beating eq}, the uncoupled pair of equations
\refstepcounter{equation}
$$
\label{beating stability decomposed}
\frac{\mathrm{d}a_{\parallel}}{\mathrm{d}T} = (\alpha+\beta)|\bar{A}|^2\left(a_{\parallel}+e^{2i\vartheta}a_{\parallel}^*\right),\quad 
\frac{\mathrm{d}a_{\perp}}{\mathrm{d}T}=\alpha|\bar{A}|^2\left(e^{2i\vartheta}a_{\perp}^*-a_{\perp}\right),
\eqno{(\theequation \mathrm{a},\mathrm{b})}
$$
wherein $|\bar{A}|$ is provided by (\ref{beating sol}a). 

We first consider parallel perturbations. Separating (\ref{beating stability decomposed}a) into its real and imaginary parts, we find the two-dimensional real system
\begin{equation}\label{beating stability parallel system}
\dfrac{\mathrm{d}}{\mathrm{d}T}\left(\begin{array}{c}{a_{\parallel}}_r \\{a_{\parallel}}_i\end{array}\right)=|\bar{A}|^2\left(\begin{array}{cc}\zeta_r(1+\cos 2\vartheta)-\zeta_i\sin2\vartheta & \zeta_r\sin2\vartheta-\zeta_i(1-\cos 2\vartheta) \\\zeta_r\sin2\vartheta+\zeta_i(1+\cos 2\vartheta) & \zeta_r(1-\cos 2\vartheta)+\zeta_i\sin2\vartheta \end{array}\right)\left(\begin{array}{c}{a_{\parallel}}_r\\{a_{\parallel}}_i\end{array}\right),
\end{equation}
wherein $\zeta=\alpha+\beta$. Seeking solutions in the form $\{{a_\parallel}_r,{a_\parallel}_i\}=\{\tilde{a}{{}_{\parallel}}_r,\tilde{a}{{}_{\parallel}}_i\}\exp(\sigma_{\parallel}T) + \text{c.c.}$, we find the growth-rate eigenvalues
\refstepcounter{equation}
$$
\label{beating parallel eigenvalues}
\sigma_{\parallel}\ub{1} = 0, \quad \sigma_{\parallel}\ub{2}=-2\gamma_r,
\eqno{(\theequation \mathrm{a},\mathrm{b})}
$$
where we have used (\ref{beating sol}a) for $|\bar{A}|$. The zero eigenvalue $\sigma_{\parallel}\ub{1}$ reflects the arbitrariness of the phase $\vartheta$. Since $\gamma_r>0$, the eigenvalue $\sigma_{\parallel}\ub{2}$ has a negative real part. Thus, planar beating is stable to parallel perturbations apart from the expected neutrality with respect to phase. 

We next consider perpendicular perturbations. Separating (\ref{beating stability decomposed}b) into its real and imaginary parts, we find the two-dimensional real system
\begin{equation}\label{beating stability perpendicular system}
\dfrac{\mathrm{d}}{\mathrm{d}T}\left(\begin{array}{c}{a_{\perp}}_r \\ {a_{\perp}}_i\end{array}\right)=|\bar{A}|^2\left(\begin{array}{cc}\alpha_r(\cos 2\vartheta-1)-\alpha_i\sin2\vartheta & \alpha_r\sin2\vartheta+\alpha_i(1+\cos 2\vartheta) \\\alpha_r\sin2\vartheta+\alpha_i(\cos 2\vartheta-1) & -\alpha_r(1+\cos 2\vartheta)+\alpha_i\sin2\vartheta \end{array}\right)\left(\begin{array}{c}{a_{\perp}}_r \\ {a_{\perp}}_i\end{array}\right).
\end{equation}
Seeking solutions in the form $\{{a_\perp}_r,{a_\perp}_i\}=\{\tilde{a}{{}_{\perp}}_r,\tilde{a}{{}_{\perp}}_i\}\exp(\sigma_{\perp}T) + \text{c.c.}$, we readily find the growth-rate eigenvalues
\refstepcounter{equation}
$$
\label{beating perpendicular eigenvalues}
\sigma_{\perp}\ub{1} = 0, \quad \sigma_{\perp}\ub{2}=\frac{2\alpha_r\gamma_r}{\alpha_r+\beta_r},
\eqno{(\theequation \mathrm{a},\mathrm{b})}
$$
where we have used (\ref{beating sol}a) for $|\bar{A}|$. The zero eigenvalue $\sigma_{\perp}\ub{1}$ reflects the arbitrariness of the direction $\be$. Since $\gamma_r>0$, whereas $\alpha_r,\beta_r<0$, the eigenvalue $\sigma_{\perp}\ub{2}$ has a positive real part. Thus, planar beating is unstable under perpendicular perturbations --- and therefore unstable under general perturbations. 

\subsubsection{Stability of the circling-whirling states}\label{sssec:cw_stability}
For the circular-whirling states (\S\ref{sssec:cw}), the form of $\bar{\bA}$ is given by \eqref{cw form}, with the plus-minus sign determining the direction of motion and the unit vector $\be$ representing a phase; without loss of generality, we only consider the stability of circular whirling in the clockwise direction, corresponding to choosing the plus sign. Writing $\ba=a_1\be+a_2\unit\times\be$, the linearised amplitude equation \eqref{linear amplitude} gives, upon eliminating $\nu$ using the steady relation \eqref{cw eq}, the coupled pair of equations
\begin{subequations}
 \label{cw stability decomposed} 
 \begin{eqnarray}
\frac{\mathrm{d}a_1}{\mathrm{d}T}=\alpha|\bar{\bA}|^2(a_1+ia_2)+\frac{1}{2}\beta|\bar{\bA}|^2(a_1+a_1^*+ia_2^*-ia_2),\\
\frac{\mathrm{d}a_2}{\mathrm{d}T}=\alpha|\bar{\bA}|^2(a_2-ia_1)+\frac{1}{2}\beta|\bar{\bA}|^2(a_2-a_2^*+ia_1+ia_1^*),
\end{eqnarray}
\end{subequations}
wherein $|\bA|$ is provided by  (\ref{cw sol}a).
Decomposing \eqref{cw stability decomposed} into their real and imaginary parts, we find the four-dimensional real system
\begin{equation}
\label{cw stability perpendicular system}
\frac{\mathrm{d}}{\mathrm{d}T}\left(\begin{array}{c}{a_1}_r \\{a_1}_i \\{a_2}_r \\{a_2}_i\end{array}\right)=|\bar{\bA}|^2\left(\begin{array}{cccc} \alpha_r+\beta_r & -\alpha_i & -\alpha_i & \beta_r-\alpha_r \\\alpha_i+\beta_i & \alpha_r & \alpha_r & \beta_i-\alpha_i \\ \alpha_i-\beta_i & \alpha_r & \alpha_r & -(\alpha_i+\beta_i) \\ \beta_r-\alpha_r & \alpha_i & \alpha_i & \alpha_r+\beta_r\end{array}\right)\left(\begin{array}{c}{a_1}_r \\{a_1}_i \\{a_2}_r \\{a_2}_i\end{array}\right).
\end{equation}
Seeking solutions in the form $\{{a_1}_r,{a_1}_i,{a_2}_r,{a_2}_i\}=\{\tilde{a}{{}_{1}}_r,\tilde{a}{{}_{1}}_i,\tilde{a}{{}_{2}}_r,\tilde{a}{{}_{2}}_i\}\exp(\sigma T) + \text{c.c.}$, we find the dispersion relation 
\begin{equation}\label{cw dispersion}
\sigma(\sigma-2|\bar{\bA}|^2\beta_r)\left(\sigma^2-4\alpha_r|\bar{\bA}|^2\sigma+4|\bar{\bA}|^4|\alpha|^2\right)=0,
\end{equation}
whose solutions yield the growth-rate eigenvalues
\refstepcounter{equation}\label{cw eigenvalues}
$$
\sigma\ub{1}=0, \quad \sigma\ub{2}=-2\gamma_r, \quad 
\sigma\ub{3}= - 2\frac{\gamma_r}{\beta_r}\alpha, \quad \sigma\ub{4}= - 2\frac{\gamma_r}{\beta_r}\alpha^*,
\eqno{(\theequation \mathrm{a}\!-\!\mathrm{d})}
$$
where we have used (\ref{cw sol}a) for $|\bar{\bA}|$. The zero eigenvalue $\sigma\ub{1}$ reflects the arbitrariness of the phase unit-vector $\be$. Since $\gamma_r>0$, whereas $\alpha_r,\beta_r<0$, the remaining eigenvalues have negative real parts. Thus, the circular-whirling states are stable apart from the expected neutrality with respect to phase. 

The predicted stability characteristics of the different states are depicted in the bifurcation diagram Fig.~\ref{fig:states}(a). The stability of the circular-whirling states and the instability of the planar-beating states (under general three-dimensional perturbations) are in accordance with the Floquet analyses numerically performed by \citet{Clarke:24}. As first established by \citet{Clarke:24}, these stability characteristics serve to rationalise the selection of whirling over beating in initial-value simulations not restricted to planar deformations \cite{Ling:18}. 

\subsection{Nonlinear stability of the undeformed state}\label{ssec:nonlinear_stab}
It remains to determine the nonlinear stability of the undeformed state at the linear-instability threshold. Setting $\chi=0$, the amplitude equation \eqref{amplitude} reduces to 
\begin{equation}
\label{amplitude threshold}
\frac{\mathrm{d}\bA}{\mathrm{d}T}=\alpha\bA^*\bA\bcdot\bA+\beta\bA\bA^*\bcdot\bA.
\end{equation}
[In the case $\chi=0$, the small parameter $\epsilon$ simply measures the magnitude of the perturbation, which scales with the $1/2$ power of that parameter.] We wish to determine whether $|\bA|$ decays or grows as $T\to\infty$. To this end, we set $\chi=0$ in (\ref{relations}a,c) to find the pair of equations
\refstepcounter{equation}
$$
\label{nonlinear stability AB eqs}
\frac{\mathrm{d}}{\mathrm{d}T}|\bA|=(\alpha_r+\beta_r)|\bA|^3-\alpha_r|\bB|^2|\bA|^{-1}, \quad \frac{\mathrm{d}}{\mathrm{d}T}|\bB|=2\beta_r|\bA|^2|\bB|,
\eqno{(\theequation \mathrm{a},\mathrm{b})}
$$
where we denote $\bB=\bA^*\times\bA$. 

In the special case of planar perturbations, where $\bB$ vanishes, (\ref{nonlinear stability AB eqs}a) reduces to 
\begin{equation}\label{nonlinear stability planar}
\frac{\mathrm{d}|\bA|}{\mathrm{d}T}=(\alpha_r+\beta_r)|\bA|^3.
\end{equation}
Since $\alpha_r,\beta_r<0$, planar perturbations decay. Explicitly, integration of \eqref{nonlinear stability planar} gives  
\begin{equation}\label{nonlinear stability planar sol}
\left|\bA(T)\right|= \frac{|\bA(0)|}{\sqrt{1-2|\bA(0)|^{2}(\alpha_r+\beta_r)T}}\sim \frac{1}{\sqrt{-2(\alpha_r+\beta_r)T}}\quad \text{as} \quad T\to\infty,
\end{equation}
revealing algebraic decay with the $-1/2$ power of time.

For initially non-planar perturbations we consider the pair of equations \eqref{nonlinear stability AB eqs} as a nonlinear dynamical system for the magnitudes $|\bA|$ and $|\bB|$. This system has no fixed points, and consideration of the phase space shows that all orbits in the first quadrant ($|\bA|,|\bB|>0$) approach the origin. Positing a power-law behaviour for $|\bA|$ and $ |\bB|$, we readily identify the late-time behaviours 
\refstepcounter{equation}
$$
\label{nonplanar powerlaws}
|\bA(T)|\sim \frac{1}{\sqrt{-2\beta_rT}}, \quad |\bB(T)|\sim -\frac{1}{2\beta_rT} \quad \text{at} \quad T\to\infty. 
\eqno{(\theequation \mathrm{a},\mathrm{b})}
$$
Given these asymptotes, we further find using \eqref{axes} that the semi-major and semi-minor axes of the fast-scale tip orbit behave as (up to an $\epsilon^{1/2}$ scaling) 
\begin{equation}
\sqrt{|\bA|^2+|\bB|}\pm \sqrt{|\bA|^2-|\bB|} = \frac{1}{\sqrt{-\beta_rT}} + o(T^{-1/2}) \quad \text{as} \quad T\to\infty.
\end{equation} 
Thus, at late times the filament exhibits circular whirling on the fast scale, with the tip radius decaying algebraically with the $-1/2$ power of time. 
\begin{figure}[t]
\begin{center}
\includegraphics[scale=0.5,trim={0.5cm 1cm 0 0}]{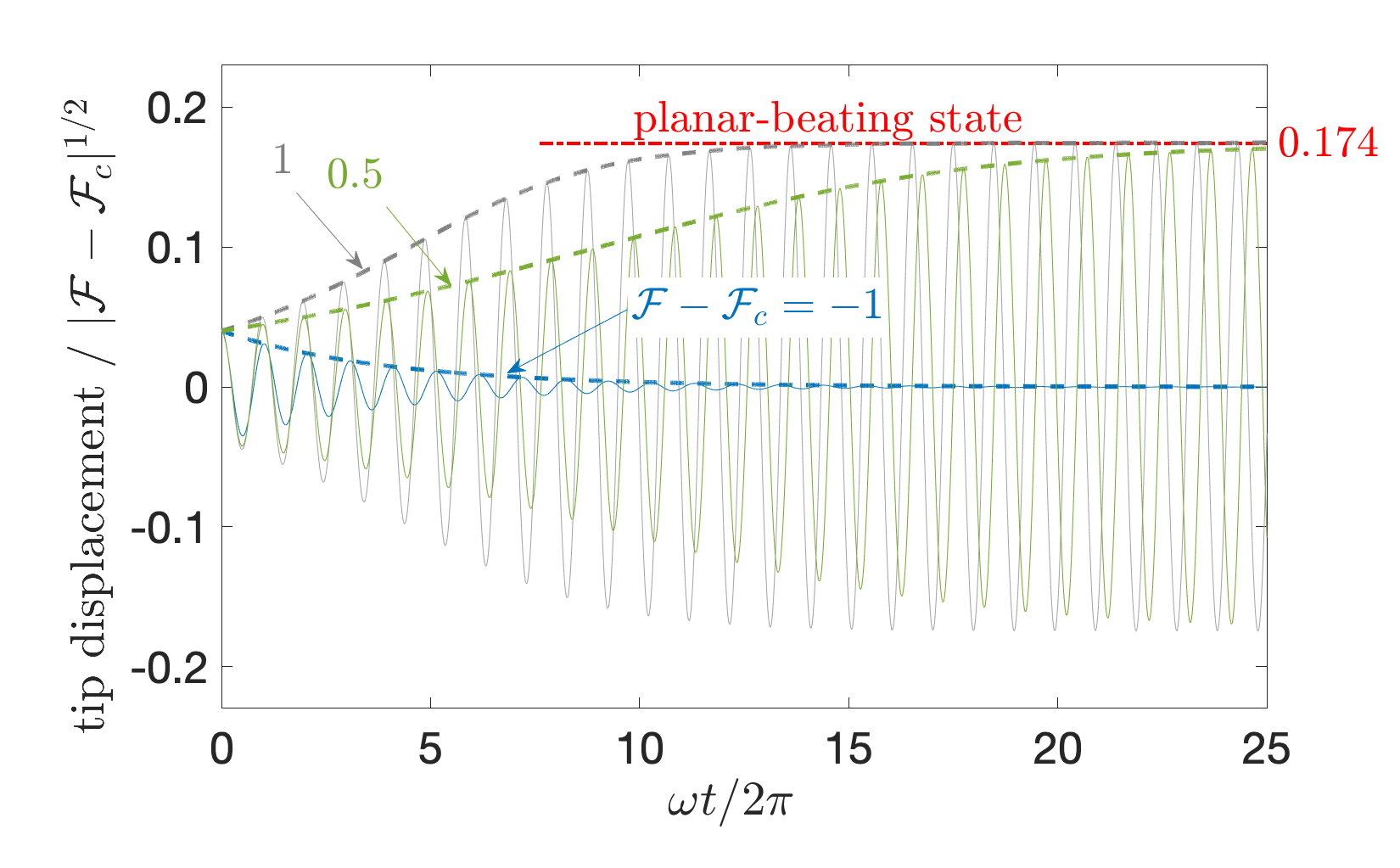}
\caption{Planar multiple-scale dynamics for the initial condition $\bA(0)=0.02\be_x$ and indicated values of $\mathcal{F}-\mathcal{F}_c$, where $\be_x$ is a unit vector parallel to the wall.  
The plot depicts the tip displacement in the $\be_x$ direction (solid curves), scaled by $|\mathcal{F}-\mathcal{F}_c|^{1/2}$, as a function of time, scaled by the natural filament period $2\pi/\omega$. The dashed curves depict the slow-time envelopes $2|\bA(T)|$. The dash-dotted line marks the peak displacement corresponding to the planar-beating state.}
\label{fig:planar}
\end{center}
\end{figure}
\begin{figure}[t!]
\begin{center}
\includegraphics[scale=0.69,trim={4cm 2cm 4cm 1cm}]{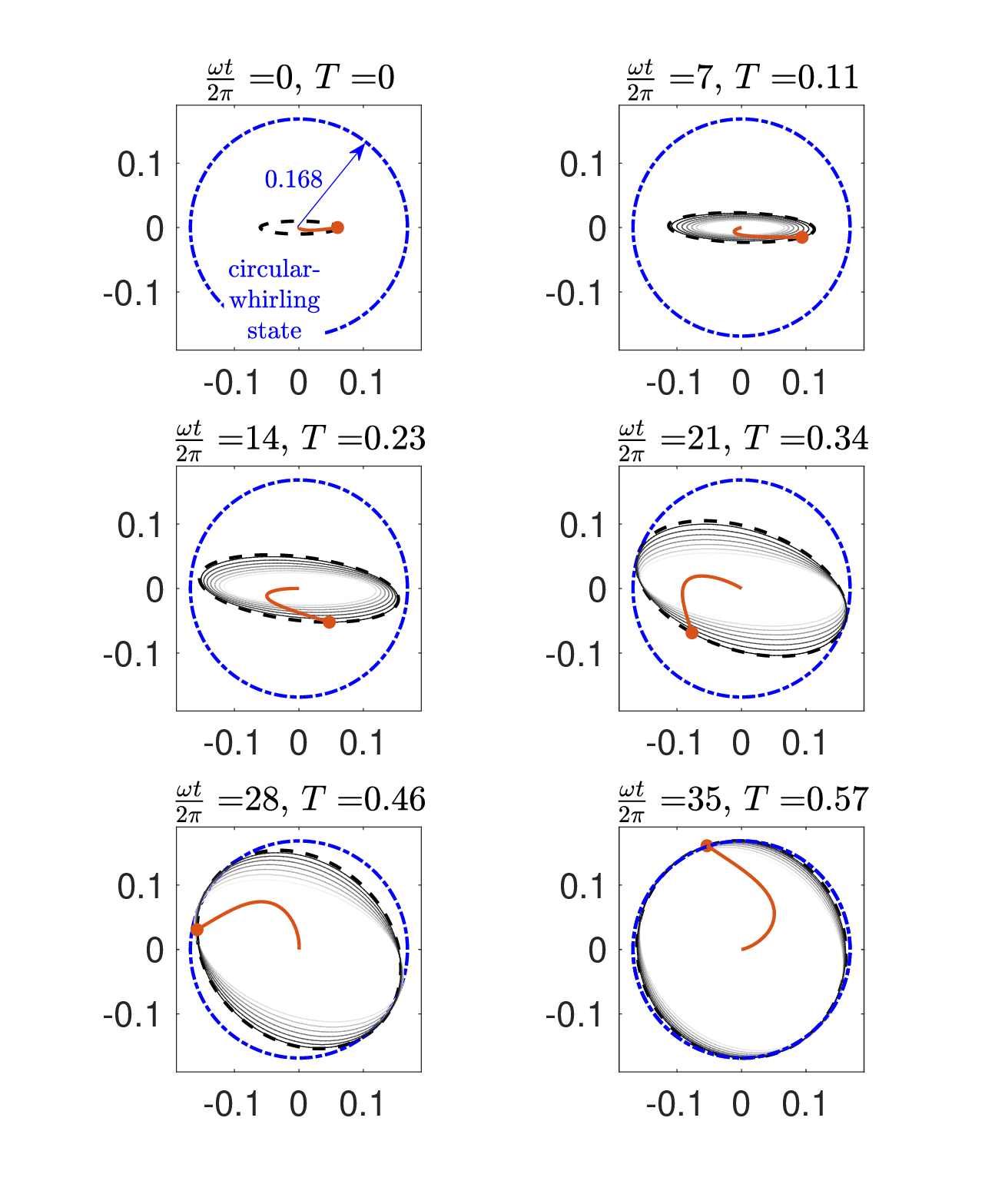}
\caption{Non-planar multiple-scale dynamics for the initial condition $\bA(0)=0.03\be_x+i0.005\be_y$ and $\mathcal{F}-\mathcal{F}_c=0.5$, where $\{\be_x,\be_y,\unit\}$ is a right-handed system of unit vectors. The subplots depict a top view (looking towards the wall, with $\be_x$ pointing to the right) showing the tip position (filled circle) and filament projection (thick solid line) at the indicated times, along with the instantaneous fast-scae tip orbit (dashed ellipse), tip trajectory starting from the previous time stamp (fading thin curves) and the radius corresponding to the circular-whirling states (dash-dotted circle); distances are scaled by $\sqrt{\mathcal{F}-\mathcal{F}_c}$.}
\label{fig:nearplanar}
\end{center}
\end{figure}
\begin{figure}[t]
\begin{center}
\includegraphics[scale=0.45,trim={2cm 1.5cm 0 0}]{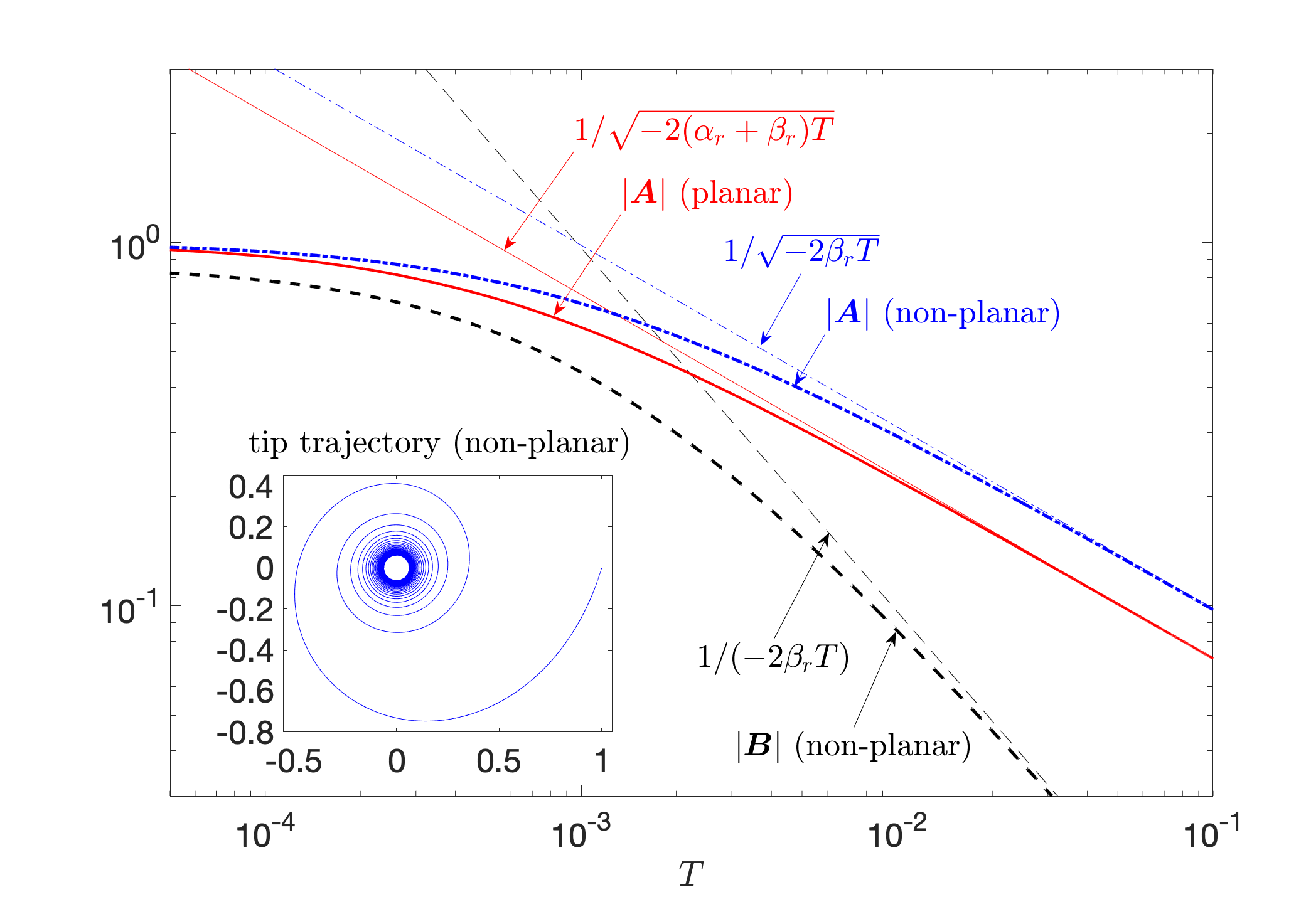}
\caption{Algebraic decay of perturbations at the linear-instability threshold $\mathcal{F}=\mathcal{F}_c$ (see \S\ref{ssec:nonlinear_stab}). The plot compares $|\bA(T)|$ in the planar scenario $\bA(0)=\be_x$ (thick solid curve) and non-planar scenario $\bA(0)=(\be_x+i\sqrt{3}\be_y)/2$ (thick dash-dotted curve) with the power-law asymptotes \eqref{nonlinear stability planar sol} and (\ref{nonplanar powerlaws}a), where $\{\be_x,\be_y,\unit\}$ is a right-handed system of unit vectors. 
For the non-planar scenario, we also depict the magnitude $|\bB|=|\bA^*\times\bA|$ (thick dashed curve) and its power-law asymptote (\ref{nonplanar powerlaws}b), and in the inset a top view (looking towards the wall, with $\be_x$ pointing to the right) of the tip trajectory scaled by $\epsilon^{1/2}$, with $\epsilon=0.5$.}
\label{fig:threshold}
\end{center}
\end{figure}
\subsection{Numerical illustrations}\label{ssec:illustrate}
The amplitude equation \eqref{amplitude} is straightforward to integrate numerically starting from some initial amplitude $\bA(0)$. Given a solution $\bA(T)$, a function of the slow time $T$, the corresponding motion of the filament in three dimensions as well as the internal force and moment distributions can be calculated at any time $t$ by making the substitution $\tilde{\bA}\Rightarrow \epsilon^{1/2}\bA(T)$ in \eqref{general linear} [cf.~\eqref{rescalings}]. For the purpose of illustrating the theory, we present in Figs.~\ref{fig:planar}-\ref{fig:threshold} sample solutions for several values of $\mathcal{F}-\mathcal{F}_c=\epsilon\chi$ and initial conditions $\bA(0)$. 

In Fig.~\ref{fig:planar}, we consider the case where the initial fast-time oscillation is planar. In light of (\ref{relations}a), the motion of the filament remains planar at all times. We observe the multiple-scales evolution of the dynamics towards the periodic planar-beating states (see \S\ref{sssec:beating} and \S\ref{sssec:beating_stability}), for $\mathcal{F}>\mathcal{F}_c$, or the undeformed configuration, for $\mathcal{F}<\mathcal{F}_c$, with faster transients for larger $|\mathcal{F}-\mathcal{F}_c|$. 
In Fig.~\ref{fig:nearplanar}, we consider an example where the initial fast-time oscillation is \emph{nearly}  planar and $\mathcal{F}>\mathcal{F}_c$. We observe the slow-time expansion and rotation of the fast-scale elliptical orbit of the tip as it approaches the circular orbit corresponding to the periodic circular-whirling states (see \S\ref{sssec:cw} and \S\ref{sssec:cw_stability}). Lastly, in Fig.~\ref{fig:threshold}, we demonstrate the algebraic decay of both planar and non-planar perturbations at the instability threshold, $\mathcal{F}=\mathcal{F}_c$ (see \S\ref{ssec:nonlinear_stab}). 

\section{Weakly nonlinear analysis}\label{sec:wna}
\subsection{Multiple-scale expansions near the instability threshold}
In this section, we derive the amplitude equation \eqref{amplitude}, which together with the general representation \eqref{general linear} for the linear approximation at the instability threshold constitutes the weakly nonlinear theory that we have presented, analysed and illustrated in the preceding section, \S\ref{sec:wnt}. The derivation consists of a weakly nonlinear analysis in the multiple-scales near-threshold regime identified in \S\ref{ssec:scalings}.

We write $\mathcal{F}=\mathcal{F}_c+\epsilon\chi$ for the dimensionless follower force, as in \eqref{F shift}, and consider the long-time dynamics of small-magnitude oscillations for $0<\epsilon\ll1$, holding $\chi$ fixed. 
Following the method of multiple scales \cite{Hinch:91}, we introduce the two-scale extension $\underline{\br}(s,\tau,T)$ of the centreline position vector ${\br}(s,t)$, wherein the `fast time' $\tau$ and `slow time' $T$ are treated as independent  coordinates, such that $\underline{\br}(s,\tau,T)=\br(s,t)$ on the `physical diagonal' 
\begin{equation}\label{physical diagonal}
\tau=t \quad \text{and} \quad T=\epsilon t;
\end{equation}
we also introduce analogous extensions $\underline{\bt}(s,\tau,T)$, $\underline{\bF}(s,\tau,T)$ and $\underline{\bM}(s,\tau,T)$ for the tangent unit vector $\bt(s,t)$, internal force $\bF(s,t)$ and internal moment $\bM(s,t)$, respectively. The extended fields satisfy the same problem as formulated in \S\ref{sec:formulation}, only with the time derivative in (\ref{eqs}b) transformed according to 
\begin{equation}\label{time trans}
\pd{}{t}\Rightarrow \pd{}{\tau}+\epsilon\pd{}{T}.
\end{equation}
We posit the `weakly nonlinear expansion'
\begin{equation}\label{rbar expansion}
\underline{\br}(s,\tau,T)={\br}_0(s)+\epsilon^{1/2}{\br}_{1/2}(s,\tau,T)+\epsilon{\br}_1(s,\tau,T)+\epsilon^{3/2}\br_{3/2}(s,\tau,T)+\cdots,
\end{equation}
with analogous expansions defined for $\underline{\bt}(s,\tau,T)$, $\underline{\bF}(s,\tau,T)$ and $\underline{\bM}(s,\tau,T)$, the zeroth-order fields being the base state \eqref{base state} evaluated \emph{at the threshold},
\begin{equation} \label{base state threshold}
\br_0=s\unit, \quad \bt_0 = \unit, \quad \bF_0=-\mathcal{F}_c\unit, \quad \bM_0=\bzero. 
\end{equation}
A key idea underlying the method is to exploit the added freedom associated with the extension to two time scales in order to ensure that the weakly nonlinear expansions remain asymptotically ordered. We note that the scaling of time by $1/\epsilon$ and the expansion in half-powers of $\epsilon$ are suggested by the scaling arguments given in \S\ref{ssec:scalings}. 
\subsection{$O(\epsilon^{1/2})$ problem} 
\label{ssec:linear_problem}
At $O(\epsilon^{1/2})$, the governing partial differential equations  \eqref{eqs}, together with the time-derivative transformation \eqref{time trans}, give 
\begin{subequations}
\label{eqs 1/2}
\begin{gather}
\pd{\br_{1/2}}{s}-\boldsymbol{t}_{1/2}=\bzero,\\
\pd{\bF_{1/2}}{s}-\left(\tI-\frac{1}{2}\unit\unit\right)\bcdot \pd{\br_{1/2}}{\tau}=\bzero, \\ 
\pd{\bM_{1/2}}{s}+\mathcal{F}_c\unit\times\boldsymbol{t}_{1/2}+\unit\times\bF_{1/2}=\bzero, \\ 
\bM_{1/2}-\unit\times\pd{\boldsymbol{t}_{1/2}}{s}=\bzero;
\end{gather}
\end{subequations}
the inextensibility constraint \eqref{constraint} gives
\begin{equation}\label{constraint 1/2}
\unit\bcdot\boldsymbol{t}_{1/2}=0;
\end{equation}
while the boundary conditions \eqref{bcs wall} and \eqref{bcs tip} give
\refstepcounter{equation}
$$
\label{bcs wall 1/2}
\br_{1/2}=\bzero, \quad \boldsymbol{t}_{1/2}=\bzero \quad \text{at} \quad s=0 
\eqno{(\theequation \mathrm{a},\mathrm{b})}
$$
and
\refstepcounter{equation}
\label{bcs tip 1/2}
$$
\bF_{1/2}+\mathcal{F}_c\boldsymbol{t}_{1/2}=\bzero, \quad \bM_{1/2}=\bzero \quad \text{at} \quad s=1. 
\eqno{(\theequation \mathrm{a},\mathrm{b})}
$$

The above problem is nothing but the full problem at the instability threshold linearised  for small deformations (scaled by $\epsilon^{1/2}$). We may accordingly deduce the appropriate general solution from the linear theory of \S\ref{sec:linear}, 
\begin{subequations}
\label{sol 1/2}
\begin{eqnarray}
\br_{1/2} &=& \bA(T)\varphi(s)e^{i\omega \tau}+\text{c.c.},\\
\boldsymbol{t}_{1/2} &=& \bA(T)\varphi'(s)e^{i\omega \tau}+\text{c.c.},\\
\bF_{1/2}&=& -\bA(T)[\varphi'''(s)+\mathcal{F}_c\varphi'(s)]e^{i\omega \tau}+\text{c.c.},\\
\bM_{1/2}&=& \unit\times\bA(T)\varphi''(s)e^{i\omega \tau}+\text{c.c.},
\end{eqnarray}
\end{subequations}
wherein $\bA(T)$ is a complex-vector amplitude parallel to the wall evolving on the slow time scale. As detailed in \S\ref{sec:linear}, this solution form represents an arbitrary superposition of linearly neutral planar-beating modes, which generally gives rise to a three-dimensional whirling motion. We have not included in the general solution \eqref{sol 1/2} potential contributions of any stable linear modes as these would exponentially decay on the fast time scale with no effect on the slow-time dynamics. We must proceed to higher order in $\epsilon$ to derive an evolution equation for $\bA(T)$. 

\subsection{$O(\epsilon)$ problem}
\label{ssec:quadratic_problem}
At $O(\epsilon)$, the governing partial differential equations \eqref{eqs}, together with the time-derivative transformation \eqref{time trans}, give
\begin{subequations}
\label{eqs 1}
\begin{gather}
\pd{\br_1}{s}-\boldsymbol{t}_1=\bzero,
\end{gather}
\begin{multline}
\pd{\bF_1}{s}-\left(\tI-\frac{1}{2}\unit\unit\right)\bcdot \pd{\br_1}{\tau} = -\frac{1}{2}\left(\boldsymbol{t}_{1/2}\unit+\unit\boldsymbol{t}_{1/2}\right)\bcdot\pd{\br_{1/2}}{\tau}\\
\left\{ =-\unit\frac{i\omega}{2}\left(\bA\bcdot\bA\varphi'\varphi e^{2i\omega\tau}-|\bA|^2\varphi'\varphi^*\right)+\text{c.c.}\right\},
\end{multline}
\begin{gather}
\pd{\bM_1}{s}+\mathcal{F}_c\unit\times\boldsymbol{t}_1+\unit\times\bF_1 = -\boldsymbol{t}_{1/2}\times\bF_{1/2} \quad 
\left\{ = \bA\times\bA^*\varphi'{\varphi^*}'''+\text{c.c.} \right\},\\
\bM_1-\unit\times\pd{\boldsymbol{t}_1}{s}=\boldsymbol{t}_{1/2}\times\pd{\boldsymbol{t}_{1/2}}{s} \quad \left\{=\bA\times\bA^*\varphi'{\varphi^*}'' + \text{c.c.}\right\};
\end{gather}
\end{subequations}
the inextensibility constraint \eqref{constraint} gives
\begin{equation}\label{constraint 1}
\unit\bcdot\boldsymbol{t}_1 = -\frac{1}{2}\boldsymbol{t}_{1/2}\bcdot\boldsymbol{t}_{1/2} \quad \left\{= -\frac{1}{2}\bA\bcdot\bA {\varphi'}^2e^{2i\omega\tau}-\frac{1}{2}|\bA|^2|\varphi'|^2 + \text{c.c.}\right\};
\end{equation}
while the boundary conditions \eqref{bcs wall} and \eqref{bcs tip} give
\refstepcounter{equation}
$$
\label{bcs surface 1}
\br_{1}=\bzero, \quad \boldsymbol{t}_1=\bzero \quad \text{at} \quad s=0
\eqno{(\theequation \mathrm{a},\mathrm{b})}
$$
and
\refstepcounter{equation}
$$
\label{bcs tip 1}
\bF_1+\mathcal{F}_c\boldsymbol{t}_1=-\chi\unit \quad \left\{=-\frac{1}{2}\chi\unit+\text{c.c.}\right\}, \quad \bM_1=\bzero \quad \text{at} \quad s=1.
\eqno{(\theequation \mathrm{a},\mathrm{b})}
$$

The homogeneous part of the above linear problem is the same as for the $O(\epsilon^{1/2})$ problem. The inhomogeneous terms, calculated by substituting \eqref{sol 1/2}, are provided explicitly inside the curly brackets; they are normal to the wall and composed of contributions that are harmonic in the fast time $\tau$ with angular frequencies $0$ or $2\omega$. Accordingly, these forcing terms cannot  resonantly excite homogeneous beating/whirling solutions of the form \eqref{sol 1/2}, which are parallel to the wall and harmonic in the fast time $\tau$ with angular frequency $\omega$. Indeed, it is straightforward to derive the solutions 
\begin{subequations}
\label{sol 1}
\begin{gather}
\br_1=-\unit\frac{1}{2}\bA\bcdot\bA e^{2i\omega\tau}\int_0^s \{{\varphi'}(p)\}^2\,dp-\unit\frac{1}{2}|\bA|^2\int_0^s|\varphi'(p)|^2\,dp + \text{c.c.},\\
\boldsymbol{t}_1=-\unit\frac{1}{2}\bA\bcdot\bA {\varphi'}^2e^{2i\omega\tau}-\unit\frac{1}{2}|\bA|^2|\varphi'|^2 + \text{c.c.},
\end{gather}
\begin{multline}
\bF_1=\frac{1}{2}\unit \bA\bcdot\bA e^{2i\omega\tau}\left[\mathcal{F}_c\{\varphi'(1)\}^2+\frac{i\omega}{2}\varphi^2(1)-\frac{i\omega}{2}\varphi^2(s)+i\omega\int_s^1dp\,\int_0^p dq\,\{\varphi'(q)\}^2\right]\\
-\unit\frac{\chi}{2}+\unit\frac{\mathcal{F}_c}{2}|\bA|^2|\varphi'(1)|^2-\unit\frac{i\omega}{2}|\bA|^2\int_s^1\varphi'(p)\varphi^*(p)\,dp +\text{c.c.}
\end{multline}
\begin{equation}
\bM_1=\bA\times\bA^*\varphi'{\varphi^*}'' + \text{c.c.},
\end{equation}
\end{subequations}
whose periodicity in $\tau$ confirms the regularity of the weakly nonlinear expansions to $O(\epsilon)$ [cf.~\eqref{rbar expansion}]. In \eqref{sol 1}, we have discarded all homogeneous solutions, including beating/whirling solutions of the form \eqref{sol 1/2} as well as solutions decaying on the fast time scale ---  these do not effect the slow-time dynamics of the leading-order amplitude $\bA(T)$.

\subsection{$O(\epsilon^{3/2})$ problem}
\label{ssec:cubic_problem}
At $O(\epsilon^{3/2})$, the governing partial differential equations \eqref{eqs}, together with the time-derivative transformation \eqref{time trans}, give
\begin{subequations}
\label{eqs 3/2}
\begin{equation}
\pd{\br_{3/2}}{s}-\boldsymbol{t}_{3/2}=\bzero,
\end{equation}
\begin{multline}
\quad \pd{\bF_{3/2}}{s}-\left(\tI-\frac{1}{2}\unit\unit\right)\bcdot \pd{\br_{3/2}}{\tau} = \left(\tI-\frac{1}{2}\unit\unit\right)\bcdot \pd{\br_{1/2}}{T}-\frac{1}{2}\left(\boldsymbol{t}_{1/2}\unit+\unit\boldsymbol{t}_{1/2}\right)\bcdot\pd{\br_{1}}{\tau}\\ - \frac{1}{2}\left(\boldsymbol{t}_1\unit+\unit\boldsymbol{t}_1+\boldsymbol{t}_{1/2}\boldsymbol{t}_{1/2}\right)\bcdot\pd{\br_{1/2}}{\tau} \quad \left\{=e^{i\omega\tau}\bbf_{(2)}(s)+\text{c.c.} + \text{n.s.t.}\right\}, 
\end{multline}
\begin{multline}
\pd{\bM_{3/2}}{s}+\mathcal{F}_c\unit\times\boldsymbol{t}_{3/2}+\unit\times\bF_{3/2} = -\boldsymbol{t}_{1/2}\times\bF_{1}-\boldsymbol{t}_1\times\bF_{1/2} \\
 \left\{=e^{i\omega\tau}\bbf_{(3)}(s)+\text{c.c.} + \text{n.s.t.}\right\},
\end{multline}
\begin{equation}
\bM_{3/2}-\unit\times\pd{\boldsymbol{t}_{3/2}}{s}=\boldsymbol{t}_{1/2}\times\pd{\boldsymbol{t}_1}{s}+\boldsymbol{t}_1\times\pd{\boldsymbol{t}_{1/2}}{s} \quad \left\{=e^{i\omega\tau}\bbf_{(4)}(s)+\text{c.c.} + \text{n.s.t.}\right\};
\end{equation}
\end{subequations}
the inextensibility constraint \eqref{constraint} gives
\begin{equation}\label{constraint 3/2}
\boldsymbol{t}_{3/2}\bcdot\unit=-\boldsymbol{t}_1\bcdot\boldsymbol{t}_{1/2},
\end{equation}
where the right-hand side vanishes since $\boldsymbol{t}_1$ and $\boldsymbol{t}_{1/2}$ are perpendicular; 
and the boundary conditions \eqref{bcs wall} and \eqref{bcs tip} give
\refstepcounter{equation}
\label{bcs wall 3/2}
$$
\br_{3/2}=\bzero, \quad \boldsymbol{t}_{3/2}=\bzero \quad \text{at} \quad s=0
\eqno{(\theequation \mathrm{a},\mathrm{b})}
$$
and
\refstepcounter{equation}
$$
\label{bcs tip 3/2}
\bF_{3/2}+\mathcal{F}_c\boldsymbol{t}_{3/2}=-\chi\boldsymbol{t}_{1/2} \quad \left\{=e^{i\omega\tau}\bh_{(1)}+\text{c.c.} + \text{n.s.t.}\right\}, \quad \bM_{3/2}=\bzero \quad \text{at} \quad s=1.
\eqno{(\theequation \mathrm{a},\mathrm{b})}
$$

The above linear problem has the same form as the problems encountered in the preceding orders, except that at the present order the inhomogeneous forcing terms include potentially resonant contributions --- parallel to the wall and harmonic in the fast time $\tau$ with the natural angular frequency $\omega$. To focus attention on these  contributions, we express the forcing terms as shown inside the curly brackets. Therein, `$\text{n.s.t}$' stands for non-secular terms and represents all other contributions, while $\bbf_{(2)}$, $\bbf_{(3)}$ and $\bbf_{(4)}$ are the `reduced forcing terms' 
\begin{subequations}
\label{secular forcing terms} 
\begin{multline}
\bbf_{(2)}=\frac{\mathrm{d}\bA}{\mathrm{d}T}\varphi+\frac{i\omega}{2}\bA^*(\bA\bcdot\bA){\varphi^*}'\int_0^s\{\varphi'(p)\}^2\,\mathrm{d}p\\
-\frac{i\omega}{2}\left[\bA|\bA|^2\varphi'({\varphi^*}'\varphi-\varphi'{\varphi^*})+\bA^*(\bA\bcdot\bA)\varphi|\varphi'|^2\right],
\end{multline}
\begin{multline}
\bbf_{(3)}=
-\unit\times\bA\varphi'\chi+\unit\times\bA|\bA|^2\left\{\mathcal{F}_c\varphi'(|\varphi'(1)|^2-|\varphi'|^2)-|\varphi'|^2\varphi'''+\omega\varphi'\mathrm{Im}\int_s^1\varphi'(p)\varphi^*(p)\,\mathrm{d}p\right\}\\
 +\frac{1}{2}\unit\times\bA^*(\bA\bcdot\bA)\left[-{\varphi'}^2(\varphi'''+\mathcal{F}_c\varphi')^*+\mathcal{F}_c{\varphi^*}'\{\varphi'(1)\}^2+\frac{i\omega}{2}{\varphi^*}'\varphi^2(1) \right.\\ \left. -\frac{i\omega}{2}{\varphi^*}'\varphi^2+i\omega{\varphi^*}'\int_s^1\mathrm{d}p\,\int_0^p \mathrm{d}q\,\{\varphi'(q)\}^2\right],
\end{multline}
\begin{equation}
\bbf_{(4)}=\unit\times\bA |\bA|^2{\varphi'}^2{\varphi^*}''   + \unit\times\bA^*(\bA\bcdot\bA)\left(\varphi''|\varphi'|^2-\frac{1}{2}{\varphi'}^2{\varphi^*}''\right),
\end{equation}
\begin{equation}
\bh_{(1)}=-\chi \bA\varphi'(1),
\end{equation}
\end{subequations}
obtained by substituting \eqref{sol 1/2} and \eqref{sol 1}; their numbering will become meaningful in the following subsection. 

\subsection{Solvability condition}\label{ssec:solvability}
\subsubsection{Restricted linear problem}
\label{sssec:restricted}
The $\exp(i\omega\tau)$ Fourier component of the $O(\epsilon^{3/2})$ problem, projected parallel to the wall, furnishes a `restricted' linear problem in the form 
\refstepcounter{equation}
\label{restricted inhomogeneous}
$$
\mathsf{L}\vec{\psi} = \vec{f}, \quad \mathsf{S}\vec{\psi}(0)=\vec{g}, \quad \mathsf{T}\vec{\psi}(1)=\vec{h},
\eqno{(\theequation \mathrm{a}\!-\!\mathrm{c})}
$$
where $\vec{\psi}(s) = [ \acute{\br}(s), \,\, \acute{\boldsymbol{t}}(s), \,\, \acute{\bF}(s), \,\, \acute{\bM}(s) ]^T$ denotes the unknown column-array field, whose elements are complex vector fields parallel to the wall; $\mathcal{L}$, $\mathcal{S}$ and $\mathcal{T}$ are the matrix-differential operators 
\refstepcounter{equation}
\label{matrix operators}
$$
\mathsf{L} = \left(\begin{array}{cccc}\frac{d}{ds} & -1 & 0 & 0 \\-i\omega & 0 & \frac{d}{ds} & 0 \\0 & \mathcal{F}_c\unit\times & \unit\times & \frac{d}{ds} \\0 & -\unit\times\frac{d}{ds} & 0 & 1\end{array}\right), \quad \mathsf{S} = \left(\begin{array}{cccc}1 & 0 & 0 & 0 \\0& 1 & 0 & 0 \end{array}\right), \quad \mathsf{T} = \left(\begin{array}{cccc}0 & \mathcal{F}_c & 1 & 0 \\0& 0 & 0 & 1 \end{array}\right);
\eqno{(\theequation \mathrm{a}\!-\!\mathrm{c})}
$$
and $\vec{f}(s)=[\bbf_{(1)}(s),\,\,\bbf_{(2)}(s)\,\,\bbf_{(3)}(s),\,\,\bbf_{(4)}(s)]^T$, $\vec{g}=[\bg_{(1)},\,\,\bg_{(2)}]^T$ and $\vec{h}=[\bh_{(1)},\,\,\bh_{(2)}]^T$ are column-array forcing terms, whose elements are vectors parallel to the wall (fields for $\vec{f}(s)$, constants for $\vec{g}$ and $\vec{h}$). In the present scenario, we find from \eqref{eqs 3/2}-\eqref{bcs tip 3/2} that $\bbf_{(1)}(s)$, $\bh_{(2)}$, $\bg_{(1)}$ and $\bg_{(2)}$ vanish, whereas $\bbf_{(2)}(s)$, $\bbf_{(3)}(s)$, $\bbf_{(4)}(s)$ and $\bh_{(1)}$ are provided by \eqref{secular forcing terms}. We know from the linear theory of \S\ref{sec:linear} that the restricted problem possesses non-trivial homogeneous solutions in the form $
\vec{\psi} = \left[\ba\varphi, \,\, \ba\varphi', \,\, -\ba(\varphi'''+\mathcal{F}_c\varphi'),\,\, \unit\times\ba\varphi''\right]^T$, with $\ba$ an arbitrary complex vector parallel to the wall. We therefore expect that solutions to the inhomogeneous restricted problem exist --- whereby resonance is avoided in the $O(\epsilon^{3/2})$ problem of \S\ref{ssec:cubic_problem} --- only under certain `solvability' conditions on the forcing terms.

Below, we derive a necessary condition for existence of solutions to the restricted problem. (By alluding to the Fredholm Alternative theorem for differential operators \citep{Keener:Book}, it could be shown that the condition is also sufficient.) In deriving this solvability condition, we shall allow for the full form of the forcing terms in \eqref{restricted inhomogeneous} in order to facilitate future generalisations of the theory as discussed in \S\ref{sec:conclude}. 

\subsubsection{Adjoint operators and homogeneous solutions}
\label{sssec:adjoints}
For any pair of column-vector fields $\vec{\psi}\ub{1}(s) = [ \acute{\br}\ub{1}(s), \,\, \acute{\boldsymbol{t}}\ub{1}(s), \,\, \acute{\bF}\ub{1}(s), \,\, \acute{\bM}\ub{1}(s) ]^T$ and $\vec{\psi}\ub{2}(s) = [ \acute{\br}\ub{2}(s), \,\, \acute{\boldsymbol{t}}\ub{2}(s), \,\, \acute{\bF}\ub{2}(s), \,\, \acute{\bM}\ub{2}(s) ]^T$, we define the inner product 
\begin{equation}\label{inner product}
\langle \vec{\psi}\ub{1},\vec{\psi}\ub{2} \rangle = \int_0^1\left\{{\br^*}\ub{1}\bcdot\br\ub{2}+{\boldsymbol{t}^*}\ub{1}\bcdot\boldsymbol{t}\ub{2}+{\bF^*}\ub{1}\bcdot\bF\ub{2}+{\bM^*}\ub{1}\bcdot\bM\ub{2}\right\}\,\mathrm{d}s.
\end{equation}
Following \citep{Keener:Book}, we seek \emph{adjoint} operators $\mathsf{L}^{\dagger}$, $\mathsf{S}^{\dagger}$ and $\mathsf{T}^{\dagger}$ such that the factor
\begin{equation}
\label{J factor}
J(\vec{\psi}\ub{1},\vec{\psi}\ub{2})=\langle \mathsf{L}\vec{\psi}\ub{1},\vec{\psi}\ub{2} \rangle - \langle \vec{\psi}\ub{1}, \mathsf{L}^\dagger \vec{\psi}\ub{2}\rangle
\end{equation}
vanishes for any $\vec{\psi}\ub{1}$ satisfying the `direct boundary conditions' $\mathsf{S}\vec{\psi}\ub{1}(0)=\vec{0}$ and $\mathsf{T}\vec{\psi}\ub{1}(1)=\vec{0}$, and $\vec{\psi}\ub{2}$ satisfying the `adjoint  boundary conditions' $\mathsf{S}^\dagger\vec{\psi}\ub{2}(0)=\vec{0}$ and $\mathsf{T}^\dagger\vec{\psi}\ub{2}(1)=\vec{0}$. To this end, we integrate by parts the product 
\begin{multline}\label{product explicit}
\langle \mathsf{L}\vec{\psi}\ub{1},\vec{\psi}\ub{2} \rangle = \int_0^1 \left\{\left(\frac{d{{{\acute{\br}}}}\ub{1}{}^*}{ds}-{\acute{\boldsymbol{t}}}\ub{1}{}^*\right) \bcdot \acute{\br}\ub{2}+\left(i\omega{\acute{\br}}\ub{1}{}^*+\frac{d{\acute{\bF}}\ub{1}{}^*}{ds}\right)\bcdot\acute{\boldsymbol{t}}\ub{2} \right. \\ \left. +\left(\mathcal{F}_c\unit\times{\acute{\boldsymbol{t}}}\ub{1}{}^* + \unit\times {\acute{\bF}}\ub{1}{}^*+\frac{d{\acute{\bM}}\ub{1}{}^*}{ds}\right)\bcdot\acute{\bF}\ub{2}+\left(-\unit\times\frac{d{\acute{\boldsymbol{t}}}\ub{1}{}^*}{ds}+{\acute{\bM}}\ub{1}{}^*\right)\bcdot\acute{\bM}\ub{2}\right\}\,\mathrm{ds}
\end{multline}
and use the vector triple product to find
\begin{multline}\label{product after integration by parts}
\langle \mathsf{L}\vec{\psi}\ub{1},\vec{\psi}\ub{2} \rangle = \left[\acute{\br}\ub{2}\bcdot\acute{\br}\ub{1}{}^*+\unit\times\acute{\bM}\ub{2}\bcdot \acute{\boldsymbol{t}}\ub{1}{}^*+\acute{\boldsymbol{t}}\ub{2}\bcdot{\acute{\bF}}\ub{1}{}^*+\acute{\bF}\ub{2}\bcdot{\acute{\bM}}\ub{1}{}^*\right]_0^1 \\
+\int_0^{1}\left\{{\acute{\br}}\ub{1}{}^*\bcdot\left(-\frac{d\acute{\br}\ub{2}}{ds}+i\omega\acute{\boldsymbol{t}}\ub{2}\right)+{\acute{\boldsymbol{t}}}\ub{1}{}^*\bcdot\left(-\acute{\br}\ub{2}-\mathcal{F}_c\unit\times\acute{\bF}\ub{2}-\unit\times\frac{d\acute{\bM}\ub{2}}{ds}\right) \right. \\ \left.
+{\acute{\bF}}\ub{1}{}^*\bcdot\left(-\frac{d\acute{\boldsymbol{t}}\ub{2}}{ds}-\unit\times\acute{\bF}\ub{2}\right)+{\acute{\bM}}\ub{1}{}^*\bcdot\left(-\frac{d\acute{\bF}\ub{2}}{ds}+\acute{\bM}\ub{2}\right)\right\}\,\mathrm{d}s.
\end{multline}
Thus, by inspection, the adjoint operators are 
\refstepcounter{equation}
$$
\mathsf{L}^\dagger = \left(\begin{array}{cccc}-\frac{d}{ds} & i\omega & 0 & 0 \\-1 & 0 & -\mathcal{F}_c\unit\times  & -\unit\times\frac{d}{ds} \\0 &  -\frac{d}{ds} & -\unit\times & 0 \\0 & 0 & -\frac{d}{ds} & 1\end{array}\right), \quad \mathsf{S}^\dagger = \left(\begin{array}{cccc}0 & 1 & 0 & 0 \\0& 0 & 1 & 0 \end{array}\right), \quad 
\mathsf{T}^\dagger = \left(\begin{array}{cccc}1 & 0 & 0 & 0 \\0& -\mathcal{F}_c & 0 & \unit\times \end{array}\right).
\eqno{(\theequation \mathrm{a}\!-\!\mathrm{c})}
$$

Associated with the adjoint operators is the adjoint homogeneous problem,
\refstepcounter{equation}
\label{adjoint problem}
$$
\mathsf{L}^\dagger \vec{\zeta} = \vec{0}, \quad \mathsf{S}^\dagger\vec{\zeta}(0)=\vec{0}, \quad \mathsf{T}^\dagger\vec{\zeta}(1)=\vec{0}.
\eqno{(\theequation \mathrm{a}\!-\!\mathrm{c})}
$$
It is readily found to possess the non-trivial solutions 
\begin{equation}\label{adjoint solutions}
\vec{\zeta}=\left[\bb\phi,\,\, \bb\frac{1}{i\omega}\phi',\,\, \unit\times\bb\frac{1}{i\omega}\phi'', \,\, \unit\times\bb\frac{1}{i\omega}\phi'''\right]^T,
\end{equation}
where $\bb$ is an arbitrary complex vector parallel to the wall and the adjoint eigenfunction $\phi(s)$ satisfies the ordinary differential equation 
\begin{equation}\label{adjoint eq}
\phi''''+\mathcal{F}_c\phi''-i\omega \phi =0
\end{equation}
and the boundary conditions 
\refstepcounter{equation}
\label{adjoint bcs}
$$
\phi' \quad \phi'' =0 \quad \text{at} \quad s=0; \quad \phi=0, \quad \phi'''+\mathcal{F}_c\phi'=0 \quad \text{at} \quad s=1. 
\eqno{(\theequation \mathrm{a}\!-\!\mathrm{d})}
$$
The unidimensional problem \eqref{adjoint eq}-\eqref{adjoint bcs} defines the adjoint eigenfunction $\phi(s)$ up to an arbitrary complex prefactor, which we conveniently choose such that $\phi(0)=1$; the eigenfunction can be readily expressed analytically in terms of the numerical values $\mathcal{F}_c$ and $\omega$, or calculated numerically. 

\subsubsection{Solvability condition}
In the definition \eqref{J factor} of the factor $J(\vec{\psi}\ub{1},\vec{\psi}\ub{2})$, let $\vec{\psi}\ub{1}=\vec{\psi}$ be a solution to the inhomogeneous problem \eqref{restricted inhomogeneous} and $\vec{\psi}\ub{2}=\vec{\zeta}$ any solution to the homogeneous adjoint problem [cf.~\eqref{adjoint solutions}]. Using (\ref{restricted inhomogeneous}a) and (\ref{adjoint problem}a), we find from \eqref{J factor} the relation 
\begin{equation}\label{solvability relation implicit}
J(\vec{\psi},\vec{\zeta})=\langle \vec{f},\vec{\zeta}\rangle.
\end{equation}
Since $\vec{\psi}$ satisfies inhomogeneous conditions, the factor $J(\vec{\psi},\vec{\zeta})$ does not necessarily vanish; rather, the boundary terms in \eqref{product after integration by parts} give, upon substituting (\ref{restricted inhomogeneous}b,c) and \eqref{adjoint solutions}-\eqref{adjoint bcs},  
\begin{equation}\label{J explicit}
J(\vec{\psi},\vec{\zeta})=\bb\bcdot\left(\frac{\phi'(1)}{i\omega}\bh_{(1)}^*+\frac{\phi''(1)}{i\omega}\bh_{(2)}^*\times\unit-\phi(0)\bg_{(1)}^*+\frac{\phi'''(0)}{i\omega}\bg_{(2)}^*\right). 
\end{equation}
From (\ref{restricted inhomogeneous}a) we find, using the triple-vector-product identity, 
\begin{equation}\label{forcing explicit}
\langle \vec{f},\vec{\zeta}\rangle=\bb\bcdot\int_0^1\left(\bbf_{(1)}^*\phi+\bbf_{(2)}^*\frac{\phi'}{i\omega}+\bbf_{(3)}^*\times\unit\frac{\phi''}{i\omega}+\bbf_{(4)}^*\times\unit\frac{\phi'''}{i\omega}\right)\,\mathrm{d}
s.
\end{equation}
Substituting \eqref{J explicit} and \eqref{forcing explicit} into \eqref{solvability relation implicit}, we find, given that $\bb$ is arbitrary, 
\begin{multline}\label{general solvability}
\int_0^1\left(\bbf_{(1)}^*\phi+\bbf_{(2)}^*\frac{\phi'}{i\omega}+\bbf_{(3)}^*\times\unit\frac{\phi''}{i\omega}+\bbf_{(4)}^*\times\unit\frac{\phi'''}{i\omega}\right)\,\mathrm{d}s\\ 
=\frac{\phi'(1)}{i\omega}\bh_{(1)}^*+\frac{\phi''(1)}{i\omega}\bh_{(2)}^*\times\unit-\phi(0)\bg_{(1)}^*+\frac{\phi'''(0)}{i\omega}\bg_{(2)}^*.
\end{multline}
This result constitutes a  general solvability condition that could be applied to any inhomogeneous linear problem in the form \eqref{restricted inhomogeneous}. In the present scenario, where $\bbf_{(1)}(s)$, $\bh_{(2)}$, $\bg_{(1)}$ and $\bg_{(2)}$ vanish, \eqref{general solvability} reduces to
\begin{equation}\label{specific solvability}
\int_0^1\left(\bbf_{(2)}{\phi^*}'-\unit\times\bbf_{(3)}{\varphi^*}''-\unit\times\bbf_{(4)}{\phi^*}'''\right)\,ds = \bh_{(1)}{\phi^*}'(1),
\end{equation} 
where we have also taken the complex conjugate. 
Substituting \eqref{secular forcing terms} for $\bbf_{(2)}(s)$, $\bbf_{(3)}(s)$, $\bbf_{(4)}(s)$ and $\bh_{(1)}$, we arrive at the amplitude equation \eqref{amplitude} with the coefficients $\alpha$, $\beta$ and $\gamma$ defined by the quadratures presented in Appendix \ref{app:coefficients}.

\section{Concluding remarks}\label{sec:conclude}
We have developed a weakly nonlinear theory illuminating the onset of spontaneous beating and whirling in the follower force model. As for other weakly nonlinear analyses, it would be relatively straightforward to generalise our approach to account for weak perturbations near the instability threshold, e.g.~perturbations to the filament or its environment and interactions with other filaments and boundaries; sufficiently close to the threshold, such perturbations are, in fact, likely to modify the \emph{leading-order} dynamics via resonant interactions with the linear beating modes. Indeed, as long as the perturbation or interaction is sufficiently weak to leave the homogeneous linearised operator unaffected, a generalised weakly nonlinear analysis could directly build on the representation of the linear approximation at the threshold, in terms of a complex-vector amplitude, and the solvability condition developed in this paper. The biological inspiration for the follower force model naturally suggests several scenarios that may merit study in this manner, including cross-sectionally anisotropic bending, preferred curvature or other intrinsic broken symmetries; non-local hydrodynamic corrections and entrainment of the flow by molecular motors; and interactions with ambient flows and neighbouring filaments. 

Besides studying the onset of spontaneous dynamics under variations to the modelling, the present work could be extended in several other interesting directions. One is to carry out weakly nonlinear analyses near other critical points of the dynamics, e.g.~near the bifurcations discovered by \citet{Clarke:24} where the quasi-periodic branch of solutions termed `QP1' merges with the planar-beating and whirling states, modifying their stability. Another direction is to consider weak perturbation and interaction effects \emph{away} from critical points; such perturbations can again have a leading-order influence over long times, e.g.~manifested in slow phase variations or slow reorientation of the beating plane. Lastly, it may be of interest to consider the onset of spontaneous dynamics in the \emph{inertial} version of the follower force model, where a finite-mass filament deforms in the absence of viscous effects. The inertial problem was widely considered in the elasticity literature (see, e.g.~\citep{Beck:52,Langthjem:00}), although mostly assuming planar deformations. A three-dimensional nonlinear analysis in the spirit of this work may accordingly prove illuminating. 

\textbf{Acknowledgements.} The author is grateful to Dr Eric E.~Keaveny of Imperial College London for introducing him to this problem and  providing the numerical results presented in \S\ref{ssec:validation}. The author acknowledges the support of the Leverhulme Trust through Research Project Grant RPG-2021-161. 
 
\appendix

\section{Absence of twist}\label{app:twist}
When including resistance to twist, the constitutive equation \eqref{constitutive dim} generalises to 
\begin{equation}\label{constitutive twist}
\mathbf{M}_*=B_*\bt\times\pd{\bt}{s_*}+T_*\left(\hat{\boldsymbol{\nu}}\bcdot\pd{\hat{\boldsymbol{\mu}}}{s_*}\right)\bt,
\end{equation}
where $T_*$ is the twist stiffness and the unit-vector triplet $\{\hat{\boldsymbol{\mu}},\hat{\boldsymbol{\nu}},\bt\}$ is a right-handed orthogonal material frame \cite{Schoeller:21}. Following \citet{Landau:BookE}, we confirm below that the twist term in \eqref{constitutive twist} vanishes identically for the scenario considered in this paper. We also show how that condition could be used to calculate the material frame along the filament given its centreline, the latter independently determined by the bending problem formulated in \S\ref{sec:formulation}.  

The dot product of the equilibrium relation (\ref{FM eqs}b) and $\bt$ gives
\begin{equation}\label{twist rel a}
\bt\bcdot\pd{\bM_*}{s_*}=\bzero.
\end{equation}
Similarly, the dot product of the generalised constitutive relation \eqref{constitutive twist} and $\partial\bt/\partial s_*$ gives, 
\begin{equation}\label{twist rel b}
\pd{\bt}{s_*}\bcdot\bM_*=\bzero,
\end{equation}
since $\partial\bt/\partial s_*$ is perpendicular to $\bt$. Combining \eqref{twist rel a} and \eqref{twist rel b}, we find
\begin{equation}\label{twist rel c}
\pd{}{s_*}(\bt\bcdot\bM_*)=0,
\end{equation}
whereby integration together with the tip boundary condition (\ref{bc tip dim}b) yields
\begin{equation}
\bt\bcdot\bM_*=0.
\end{equation}
It follows that the twist term in \eqref{constitutive twist} vanishes. 

The last result implies the orthogonality relation 
\begin{equation}\label{mu_s nu}
\hat{\boldsymbol{\nu}}\bcdot\pd{\hat{\boldsymbol{\mu}}}{s_*}=0.
\end{equation}
The remaining components of $\partial\hat{\boldsymbol{\mu}}/\partial s_*$ can be found as follows. First, the constraint that $\hat{\boldsymbol{\mu}}$ is a unit vector implies  
\begin{equation}\label{mu_s mu}
\hat{\boldsymbol{\mu}}\bcdot\pd{\hat{\boldsymbol{\mu}}}{s_*}=0.
\end{equation}
Second, since $\bt$ and $\hat{\boldsymbol{\mu}}$ are orthogonal, 
\begin{equation}\label{mu_s t}
\bt\bcdot \pd{\hat{\boldsymbol{\mu}}}{s_*}=-\hat{\boldsymbol{\mu}}\bcdot\pd{\bt}{s_*}. 
\end{equation}
Combining \eqref{mu_s nu}-\eqref{mu_s t}, we find the equation
\begin{equation}
\label{mu eq}
\pd{\hat{\boldsymbol{\mu}}}{s_*}=-\left(\hat{\boldsymbol{\mu}}\bcdot\pd{\bt}{s_*}\right)\bt,
\end{equation}
which is supplemented by the boundary condition 
\begin{equation}
\label{mu bc}
\hat{\boldsymbol{\mu}}=\be \quad \text{at} \quad s_*=0,
\end{equation}
wherein $\be$ is an arbitrarily chosen unit vector parallel to the wall; since the filament is clamped at the wall, $\be$ is time-independent and can accordingly be identified with $\hat{\boldsymbol{\mu}}$ in its undeformed configuration. Given $\bt(s_*,t_*)$ from the solution to the bending problem formulated in \S\ref{sec:formulation}, \eqref{mu eq} and \eqref{mu bc} could, in principle, be solved for $\hat{\boldsymbol{\mu}}(s_*,t_*)$. The remaining unit vector $\hat{\boldsymbol{\nu}}(s_*,t_*)$ could then be obtained from $\hat{\boldsymbol{\nu}}=\bt\times \hat{\boldsymbol{\mu}}$.

\section{Coefficients appearing in the amplitude equation}
\label{app:coefficients}
The coefficients appearing in the amplitude equation \eqref{amplitude primitive}, or \eqref{amplitude}, are defined as
\begin{subequations}
\label{coefficients integrals}
\begin{multline}
\alpha = \frac{i\omega}{2\xi}\left\{\int_0^1\mathrm{d}s\,{\phi^*}'(s){\varphi^*}'(s)\int_0^s \mathrm{d}p\,\{\varphi'(p)\}^2-\int_0^1{\phi^*}'\varphi|\varphi'|^2\,\mathrm{d}s\right\} \\
+\frac{1}{2\xi}\int_0^1\mathrm{d}s\,{\phi^*}''\left\{-{\varphi'}^2(\varphi'''+\mathcal{F}_c\varphi')^*+\{\varphi'(1)\}^2\mathcal{F}_c{\varphi^*}'+\frac{i\omega}{2}\varphi^2(1){\varphi^*}' \right. \\ \left. 
-\frac{i\omega}{2}{\varphi^*}'\varphi^2+i\omega{\varphi^*}'\int_s^1 \mathrm{d}p \int_0^p \mathrm{d}q\, \{\varphi'(q)\}^2\right\}+\frac{1}{\xi}\int_0^1{\phi^*}'''(\varphi''|\varphi'|^2-\frac{1}{2}{\varphi'}^2{\varphi^*}'')\,\mathrm{d}s,
\end{multline}
\begin{multline}
\beta = -\frac{i\omega}{2\xi}\int_0^1{\phi^*}'\varphi'({\varphi^*}'\varphi-\varphi'{\varphi^*})\,\mathrm{d}s + 
\frac{1}{\xi}\int_0^1\mathrm{d}s\,{\phi^*}''\left\{-|\varphi'|^2(\varphi'''+\mathcal{F}_c\varphi')+\mathcal{F}_c|\varphi'(1)|^2\varphi'+
\right.\\\left. \frac{i\omega}{2}\varphi'\left[|\varphi(1)|^2-|\varphi|^2-2\int_s^1\mathrm{d}p\,\varphi'(p){\varphi^*}(p)\right]\right\}+\frac{1}{\xi}\int_0^1{\phi^*}'''\varphi'^2{\varphi^*}''\,\mathrm{d}s,
\end{multline}
\begin{equation}
\gamma = \frac{1}{\xi}\int_0^1{\phi^*}'\varphi''\,\mathrm{d}s,  \end{equation}
\end{subequations}
wherein $\xi = -\int_0^1{\phi^*}'\varphi\,\mathrm{d}s$. Recall that $\varphi(s)$ is the eigenfunction defined in \S\ref{sec:linear}, along with the threshold follower-force and angular-frequency values $\mathcal{F}_c$ and $\varpi$;  and $\phi(s)$ is the adjoint eigenfunction defined in \S\ref{ssec:solvability}. Since all of these quantities are parameter-free, so are the coefficients $\alpha$, $\beta$ and $\gamma$. The quadratures in \eqref{coefficients integrals} are readily performed, giving the numerical values of the coefficients quoted in \eqref{coefficients numbers}. 

\bibliography{refs}
\end{document}